\PassOptionsToPackage{dvipsnames}{xcolor}

\documentclass[article,authoryear,arxiv]{beg_32}

\usepackage[dvipsnames]{xcolor}
\usepackage{mathtools}
\usepackage{amsmath}
\usepackage{amsthm}

\usepackage{csl}
\csltitle{Improving the Adaptive Moment Estimation (ADAM) stochastic optimizer through an Implicit-Explicit (IMEX) time-stepping approach}
\cslauthor{Abhinab Bhattacharjee, Andrey A. Popov, Arash Sarshar, Adrian Sandu}
\cslyear{2024}
\cslreportnumber{2}
\cslemail{abhinab93@vt.edu}

\usepackage[listings,skins,breakable]{tcolorbox}
{%
\end{tcolorbox}\endgroup}
\newtcbox{\popovnotesmall}{breakable,enhanced jigsaw,nobeforeafter,tcbox raise base,boxrule=0.4pt,top=0mm,bottom=0mm,
  right=0mm,left=4mm,arc=1pt,boxsep=2pt,before upper={\vphantom{dlg}},
  colframe=Fuchsia!75!black,coltext=black,colback=Fuchsia!20,
  overlay={\begin{tcbclipinterior}\fill[Fuchsia!75!black] (frame.south west)
    rectangle node[text=white,font=\sffamily\bfseries\tiny,rotate=90] {AAP} ([xshift=4mm]frame.north west);\end{tcbclipinterior}}}
\MakeRobust\popovnotesmall
\ifdefined
\fi

\usepackage[hang]{footmisc}
\fancypagestyle{plain}{%
  \fancyhf{}
  \fancyhead[R]{}
  \fancyfoot[R]{}
  \fancyfoot[L]{}
  }
\renewcommand{\dmyy}{20}
\renewcommand{\myyear}{2024}
\renewcommand{\today}{}
\begin{document}

\csltitlepage

\title{Improving the Adaptive Moment Estimation
(ADAM) stochastic optimizer through an
Implicit-Explicit (IMEX) time-stepping
approach}
\titlehead{IMEX GARK}
\authorhead{Bhattacharjee, Popov, Sarshar \& Sandu}
\corrauthor[1]{Abhinab Bhattacharjee}
\author[2]{Andrey A. Popov}
\author[3]{Arash Sarshar}
\author[1]{Adrian Sandu}
\corremail{abhinab93@vt.edu}
\corraddress{Computational Science Lab, Department of Computer Science, Virginia Tech, Blacksburg, VA 24060}
\address[1]{Computational Science Lab, Department of Computer Science, Virginia Tech, Blacksburg, VA 24060}
\address[2]{The Oden Institute for Computational Engineering and Sciences, The University of Texas at Austin, TX 78712}
\address[3]{Department of Computer Engineering and Computer Science, California State University, Long Beach, CA 90840}


\abstract{The \textsc{Adam} optimizer, often used in Machine Learning for neural network training, corresponds to an underlying ordinary differential equation (ODE) in the limit of very small learning rates. 
This work shows that the classical \textsc{Adam} algorithm is a first-order implicit-explicit (IMEX) Euler discretization of the underlying ODE.
Employing the time discretization point of view, we propose new extensions of the \textsc{Adam} scheme obtained by using higher-order IMEX methods to solve the ODE.
Based on this approach, we derive a new optimization algorithm for neural network training that performs better than classical \textsc{Adam} on several regression and classification problems.
}


\maketitle

\section{Introduction}
\label{sec:introduction}


Neural network (NN) training requires solving unconstrained optimization problems ~\citep{Bengio_PracticalRecommendationforTraining} of the form
\begin{equation}\label{eq:neural-network-cost-function}
   \theta^{*} =  \argmin_\theta \mathcal{L} ([X,Y], \widehat{Y}; \theta) \quad \textnormal{subject to} \quad
   \widehat{Y} = f(X,\theta),
\end{equation}
where $[X,Y]$ are the input-output pairs of the training data set, $f$ is some parameterized NN (with $\theta$ as parameters), $\theta^*$ are the optimal parameters, $\mathcal{L}$ is some loss function, and $\widehat{Y}$ is the output of the NN for the input $X$.

Stochastic Gradient Descent (\textsc{SGD}), inspired by the early work of \citep{cauchy1847methode}, trains NNs using subsets (batches) of data to manage large datasets.
It is employed in many deep learning applications, such as regression, image classification, or sequence modeling \citep{SGD_escapes_local_minima, auto_encoding_var_bias, deep_residual_network}.
However, it often suffers from unwanted oscillations when small batches are used.
Momentum methods \citep{HeavyBall, NesterovMomentum} extend gradient descent by keeping track of the direction of persistent decrease of the loss function, which helps reduce oscillations introduced by stochastic behavior. 

First-order adaptive optimization methods use the history of gradient evaluations to improve convergence.
The adaptivity stems from the change in the learning rate according to the exponential moving average of past gradients. 
Some of the well-known adaptive methods are \textsc{ADAGrad}~ \citep{Adagard}, \textsc{RMSProp}~\citep{rmsprop}, \textsc{ADADelta}~\citep{Adadelta}, \textsc{Adam}~\citep{kingma2017Adam} and others \citep{Adaptive_Nonconvex_Optimization}.
This study focuses on \textsc{Adam}, one of the most successful and widely used optimizers in deep learning~\citep{ReddiAdamConvergence, krizhevsky2017imagenet, gpt3}. The algorithm is summarized in \ref{app:original_adam}.

It was shown in \citep{ODEAdaptiveAlgorithm} that, in the limit of small learning rates, adaptive methods approach a continuous dynamical system 
called \textit{``the underlying ODE''} \citep{ReddiAdamConvergence}. Similarly, batch training leads to underlying stochastic differential equations in the limit.

Taking advantage of the optimization problem's continuous formulation, we propose using high-order schemes to discretize the underlying ODE and derive new discrete optimization algorithms for NN training.
The discretization of stochastic differential equations, which leads to new stochastic training algorithms, has not been studied in this paper and will be addressed in future work.

The novel contributions of this work are as follows:
\begin{itemize}
    \item We interpret \textsc{Adam} as first order Generalized Additive Runge-Kutta (GARK)  IMEX Euler discretization of the underlying ODE \citep{GARKSandu}.  This formulation recovers the original algorithm proposed in \cite{kingma2017Adam}.
    \item We propose a second-order Trapezoidal IMEX method to discretize the underlying ODE. Numerical experiments show that the resulting discrete algorithm performs similar or better than standard \textsc{Adam} in a number of regression and classification problems.
\end{itemize}

The remainder of this paper is organized as follows:
\Cref{sec:background} provides the background and motivation for this study.
\Cref{sec:proposed_methodology} introduces the formulation of \textsc{Adam} as a numerical discretization of the underlying ODE. Employing high-order discretization techniques allows the development of new optimization algorithms.
\Cref{sec:numerical_experiments} reports numerical experiment results.
\Cref{sec:conclusion} concludes the study with a summary of the best method we found and discusses several future extensions. 

\section{Background}
\label{sec:background}

This work focuses on the \textsc{Adam} \citep{kingma2017Adam} numerical optimization procedure used for training deep feedforward and convolutional neural networks. The original algorithm is summarized in \Cref{alg:Adam}.

\subsection{\textsc{Adam}: an optimization method for training NNs} \label{subsec:Performance_of_Optimization}

The \textsc{Adam} method (\cref{alg:Adam}) has several tunable parameters.  $\beta$-s determine the amount of information carried from one epoch to the next or how much is exponentially forgotten from the last step/epoch. The learning rate ($h$) is another hyperparameter influencing the convergence rate. \textsc{Adam} is an ``adaptive'' learning algorithm because it scales the learning rate by the velocity ($h/\sqrt{v}$). Given the dependency on the previous states and the learning rate, it is natural to think of \textsc{Adam} as a dynamic system.

\textsc{Adam} can be rewritten as follows:
\begin{equation}
\label{eq:AdamAlternateFormulation}
 	\begin{split}
 	    g_{n+1} &= \nabla_{\theta}\mathcal{L} (t_{n+1}, \theta_{n}) ~, \\
 		m_{n+1} &= (1 - \beta_{1}) \sum\limits_{i = 0}^{n+1} \beta_{1}^i g_{n+1-i} ~, \\
 		v_{n+1} &= (1 - \beta_{2}) \sum\limits_{i = 0}^{n+1} \beta_{2}^i g_{n+1-i}^{\circ 2} ~, \\
 		\theta_{n+1} &= \theta_{n} - h  \dfrac{(1 - \beta_{1}) \sum\limits_{i = 0}^{n+1} \beta_{1}^i g_{n+1-i}}{\sqrt[\circ]{(1 - \beta_{2}) \sum\limits_{i = 0}^{n+1} \beta_{2}^i g_{n+1-i}^{\circ 2}}} ~,
 	\end{split}
\end{equation}
where the following shorthand notation is used: 
\[
\mathcal{L} (t_{n+1}, \theta) := \nabla_{\theta} \mathcal{L}([X,Y]_{n+1}, \widehat{Y}_{n+1}; \theta),
\] 
where $t_{n+1} = t_{n}+h$. 
The hyper-parameters $\beta$, and consequently, the $\alpha$'s used in \citep{ODEAdaptiveAlgorithm}, should be tuned according to the loss function. It has been shown in the literature that the original $\beta$ values perform worse than \textsc{SGD} in a contrived problem \citep{ODEAdaptiveAlgorithm, ReddiAdamConvergence}.

It has also been shown 
that, in the limit of small learning rates as $h \to 0$, the  \textsc{Adam} algorithm \cref{eq:AdamAlternateFormulation} is a numerical discretization of the following ordinary differential equation \citep{ODEAdaptiveAlgorithm}:
\begin{equation}
\label{eq:ODE}
 	\begin{split}
 		\dot{\theta}(t) &= -m(t)/\sqrt[\circ]{v(t) + \epsilon}, \\
 		\dot{m}(t) &= d(t) \, \nabla_\theta\mathcal{L}(t,\theta(t)) - r(t)\,m(t), \\
 		\dot{v}(t) &= p(t)\, [\nabla_\theta\mathcal{L}(t,\theta(t))]^{\circ 2} -q(t)\,v(t),		
 	\end{split}
\end{equation}
where $[\cdot]^{\circ 2}$ is used for element-wise square operation, $\sqrt[\circ]{\cdot}$ for element-wise square root, $[\cdot]/[\cdot]$ for element-wise division, and
``$t$'' denotes the time component in the continuous setting.
We use the short-hand notation 
%
\begin{equation}
\label{eqn:define-L}
\mathcal{L}(t,\theta) := \mathcal{L} \big([X_t,Y_t], \widehat{Y}_t; \theta\big),
\end{equation}
where $X_t,Y_t, \widehat{Y}_t$ represent the data inputs and outputs at time $t$, and model outputs at time $t$ respectively.

The $\epsilon$ in the denominator of the first equation is added for numerical stability and is fixed in this study ($\epsilon = 10^{-8}$), though there is evidence to suggest that $\epsilon$ plays a role in the converge of the algorithm \citep{Empirical_Comparisons_of_Optimizers}.
The values of the time-dependent parameters $d(t), r(t), p(t)$ and $q(t)$ for different first-order adaptive methods are given in \cite{ODEAdaptiveAlgorithm}. 
For \textsc{Adam}, the parameters are as follows:
\begin{equation}\label{eq:AdamParameter}
 	\begin{split}
 		d(t) &= r(t) = \dfrac{1}{\alpha_1} , \quad
 		p(t) = q(t) = \dfrac{1}{\alpha_2}, \\
 		\alpha_i &:= -h/\log \beta_i \quad \Leftrightarrow \quad \beta_i = e^{\frac{-h}{\alpha_i}},~i=1,2,	
 	\end{split}
\end{equation}
and depend on the learning rate $h$.

\subsection{Optimization in the Continuous and Discrete Setting} \label{subsec:Optimization_Continuous}

Solving the optimization problem in continuous formulation \cref{eq:ODE} is, in effect,  solving a system of ordinary differential equations numerically. The theory regarding the solution of ODEs~\citep{BUTCHEROverview2000, Hairer1, Hairer2} is relevant to but outside the scope of this study.
We focus our effort on ODEs of the form:
\begin{equation}
\label{eq:ODEgeneralform}
\begin{split}
    \dot{y} &= f(t,y), \quad t \in [t_0, t_f] , \quad 
    y(t_0) = y_0,
\end{split}
\end{equation}
where $f$ represents the non-linear dynamics, $y_0$ represents the initial conditions, and $t_0$ to $t_f$ represents the timespan of interest.
Two well-known numerical methods for solving  \cref{eq:ODEgeneralform} are the explicit  Forward Euler (FE) scheme 
\begin{subequations}
\label{eq:ODEExplicitandimplicit}
\begin{align}
\label{eq:ODE-FE}
    y_{n+1} &= y_{n} + h\, f(t_n, y_n),  
\end{align}
and the implicit Backward Euler (BE) scheme
\begin{align}
\label{eq:ODE-BE}
    y_{n+1} &= y_{n} + h\, f(t_{n+1}, y_{n+1}).
\end{align}
\end{subequations}
Explicit methods such as \cref{eq:ODE-FE} are computationally inexpensive per step.
However, they are conditionally stable. In contrast, implicit methods like \cref{eq:ODE-BE} generally require the solution of a (non-)linear system of equations to compute $y_{n+1}$ and are therefore computationally expensive per step. 
Implicit schemes enact a computational benefit when the ODE is stiff. “Formally, a stable system of ODEs is stiff if the eigenvalues of its Jacobian matrix have greatly differing magnitudes.” \citep{siamscientificcomputingbook}. 
In practice, the solution of stiff ODEs achieves the desired level of accuracy when solved with implicit methods with significantly larger timesteps compared to the same ODE solved with an explicit method \citep{Hairer1, Hairer2, time_integration_Ascher}.
Combining the best of both worlds, IMEX methods were developed where the right-hand side of the ODE is split into a stiff and a non-stiff part and treated with an implicit and explicit scheme respectively.
This idea is at the core of this study and is explained in more detail in \cref{subsec:IMEX_GARK}.

Following \cite{ODEAdaptiveAlgorithm}, the FE discretization of our optimization ODE in \cref{eq:ODE} is:
\begin{equation}
    \label{eq:ForwardEulerODE}
 	\begin{split}
 		\theta_{n+1} &= \theta_{n} - h \, m_{n}/\sqrt[\circ]{v_{n} + \epsilon}, \\
 		m_{n+1} &= m_{n}\, \left[1 - h \, r({t_n})\right] + h \, d({t_n})\, \nabla_\theta\mathcal{L}(t_{n},\theta_{n}), \\
 		v_{n+1} &= v_{n}\, \left[1 -h \, q({t_n})\right]  + h \, p({t_n}) \, [\nabla_\theta\mathcal{L}(t_{n},\theta_{n})]^{\circ 2}.		
 	\end{split}
\end{equation}
The discretization time step $h$ is also known as \textit{``learning rate''} in ML literature, and the two terms will often be used interchangeably in this study.
Subscripts $n$ represent the algorithmic discrete time  (often called \textit{``epoch''} in ML literature).
This notation holds for training with small batches of data, albeit with a minor change (refer to \cref{rem:assumption_batch}.)
Note the similarities and differences between the formulation in  \cref{eq:ForwardEulerODE} and the one given in \Cref{alg:Adam}.
More accurate algorithms for solving the ODE \cref{eq:ODEgeneralform} are offered by the class of Runge-Kutta (RK) methods \cite{Hairer1}.
\subsection{Linear stability of the FE discretization of the optimization ODE}\label{subsec:linear_stability_FE}
We can demonstrate the limitation of the stepsize due to the stability of the FE discretization by  considering a linearized gradient term in \cref{eq:ForwardEulerODE}:
\begin{equation}
 	\begin{split}
 		\theta_{n+1} &= \theta_{n} - h \, m_n /\sqrt{ \epsilon + v_n}, \\
 		m_{n+1} &= m_n\, \left(1 - h \, r\right) + h \, d\, \Lambda \theta_{n}, \\
 		v_{n+1} &= v_{n}\, \left(1 -h \, q\right) +  \Lambda^{\circ 2} \theta_{n}^{\circ 2},		
 	\end{split}
\end{equation}
in which the first order approximation of the gradient is considered as $\nabla_\theta\mathcal{L}(t_{n},\theta_{n}) \approx \Lambda  \theta_n$ and $p_n = p, q_n =q , d_n =d, r_n=r$ for all timesteps. Linearizing this equation and ignoring higher-order terms, we have:
\begin{equation}
    \label{eq:ForwardEulerODELinear}
    \begin{bmatrix}
        \theta \\ m \\ v
    \end{bmatrix}_{n+1} = 
    \begin{bmatrix}
        1 & -h/\sqrt{\epsilon} & 0 \\
        hd\Lambda & 1-h r & 0 \\
        0 & 0 &  1-hq
    \end{bmatrix}
    \begin{bmatrix}
        \theta \\ m \\ v
    \end{bmatrix}_{n}.
\end{equation}
Linear stability analysis (\cite{Hairer2}) requires the matrix in \cref{eq:ForwardEulerODELinear} to have stable eigenvalues, i.e, $\max \{|1-hq|, |\Re(1- \frac{hr}{2} \pm \frac{\sqrt{-4dh^2 \Lambda \sqrt{\epsilon} + h^2 r^2 \epsilon}}{2 \sqrt{\epsilon}})|\} \leq 1$. This emphasizes the fact that the stability of the FE method is conditional on step size and parameters $q$ and $r$. A stability region plot for the FE discretization  \cref{eq:ForwardEulerODE} with nominal hyperparameters is available in \cref{app:stability_plots}). Before introducing the methodology, we clarify several assumptions. 
\begin{assumptions}
    The loss function is defined as $\mathcal{L}([X,Y], \hat{Y}; \theta): \mathrm{R}^{p \times d} \to \mathrm{R}$, where $p$ and $d$ are the dimensions of the parameter and input space.
    The loss is considered to be $C^{2}$ continuous over the parameter space.
    The gradient of the loss is also considered Lipschitz continuous such that 
    \begin{equation*}
    \exists C > 0, \norm{\nabla_{\theta} \mathcal{L} ( [X, Y], \hat{Y}; \theta_{m}) - \nabla_{\theta} \mathcal{L} ([X, Y], \hat{Y}; \theta_{n})} \leq C \norm{\theta_{m} - \theta_{n}} \quad \forall (\theta_{m}, \theta_{n}) \in \mathrm{R}^{p},
    \end{equation*}
    with $([X, Y], \hat{Y})$ are the input, output data, and the predicted neural network output, respectively. This Lipschitz continuity is useful when considering noisy data, such as the case of batching, and guarantees the boundedness of the gradient. Since the gradient also appears on the right-hand side of the ODE system  \cref{eq:ODEgeneralform}, this assumption is also necessary for the existence of the numerical solution of  \cref{eq:ODEgeneralform}.
\end{assumptions}
\begin{remark}\label{rem:assumption_batch}
When there is no batching in training, the ODE \cref{eq:ODE} is continuous across all time steps(epochs.) However, using a batch of data to calculate an approximate gradient in the time interval $t_n \leq t \leq t_{n+1}$ can lead to a point-wise discontinuity at the start and end of the intervals. From a practical implementation point of view, this is not an issue for us, as can be seen in \cref{sec:numerical_experiments}.
\end{remark}
\begin{remark}\label{rem:implicit_method}
    An implicit method is unconditionally stable but not necessarily accurate. However, training a neural network may not require a very accurate approximation of the trajectory of the underlying ODE. Our aim is to reach a minima faster. If a method can do that with larger time steps, then the training is faster.
\end{remark}
\begin{remark}\label{rem:alpha_beta}
    The $\mathbf{\beta}$s from the original \textsc{Adam} \citep{kingma2017Adam} and the $\mathbf{\alpha}$s from the ODE formulation of the adaptive methods \citep{ODEAdaptiveAlgorithm} are related through the relation given by \cref{eq:AdamParameter}.
    It is important to note that the $\alpha$'s need to be calculated from the $\beta$'s using the given relationship and vice-versa
\end{remark}
%
%
    

\section{Methodology}
\label{sec:proposed_methodology}

\subsection{IMEX GARK Time Integration} \label{subsec:IMEX_GARK}

We consider now the case of ODEs of the form \cref{eq:ODEgeneralform} that can be naturally split into a non-stiff part and a stiff part:
\begin{equation}
\label{eq:ComponentSplit_IMEX}
    \begin{bmatrix} \dot{y} \\ \dot{z} \end{bmatrix} = \begin{bmatrix} f (t, y, z) \\ g (t, y, z)  \end{bmatrix}, \quad  \begin{bmatrix}  y (t_0)  \\ z(t_0) \end{bmatrix} = \begin{bmatrix} y_0 \\ z_0 \end{bmatrix},
\end{equation}
where $y$ are the non-stiff components with $f(\cdot)$ the non-stiff dynamics, and $z$ are the stiff components with $g(\cdot)$ the stiff dynamics of the system.
Solving the entire system in  \cref{eq:ComponentSplit_IMEX} with an explicit method such as FE as introduced in \cref{eq:ODE-FE} typically requires unacceptably small step sizes, due to the stability restrictions imposed by the stiff part \citep{Hairer1, Hairer2}.
Conversely, solving the entire system \cref{eq:ComponentSplit_IMEX} with an implicit method like BE in \cref{eq:ODE-BE} results in large nonlinear systems of equations that couple all the variables, both stiff and non-stiff, whose solution is computationally expensive and often intractable in higher dimensional systems. 

An intuitive idea is to solve the non-stiff part with an inexpensive explicit scheme and the stiff part with a stable implicit method.
Taking advantage of both methods gave rise to the class of Implicit-Explicit (IMEX) \citep{IMEXpaper} time discretization methods. 
For example, one can solve  \cref{eq:ComponentSplit_IMEX} as follows:
First, advance the non-stiff component from  $y_{n}$  at time $t_n$ to $y_{n+1}$  at time $t_{n+1}$ using the FE scheme given in \cref{eq:ODE-FE}.
Next, advance the stiff component from  $z_n$ at time $t_n$ to $z_{n+1}$ at  time $t_{n+1}$ using the BE scheme in \cref{eq:ODE-BE}. The resulting IMEX Euler method reads:
\begin{equation}
\label{eq:IMEXExplicit first}
\begin{bmatrix}   y_{n+1} \\ z_{n+1} \end{bmatrix} = \begin{bmatrix}  y_{n} + h\, f (t_{n}, y_{n}, z_{n}) \\ z_{n} + h\, g (t_{n+1}, y_{n+1}, z_{n+1}) \end{bmatrix}.
\end{equation}
Note that in \cref{eq:IMEXExplicit first} we first solve the explicit component, i.e., the first equation, and then use $y_{n+1}$ in the solution of the second equation.
As a result, the nonlinear system needs to be solved only for the stiff components $z_{n+1}$.

This paper focuses on the application of IMEX methods to solve the optimization ODE in \cref{eq:ODE}.
We consider Generalized-structure Additive Runge-Kutta (GARK) family of methods \citep{GARKSandu}, which are a generalized formulation of earlier partitioned Runge-Kutta schemes \citep{ARKmethods}.

A two-way partitioned (diagonally implicit) GARK method with stepsize $h$ solves the component partitioned system \cref{eq:ComponentSplit_IMEX} as follows \citep{GARKSandu}:
\begin{subequations}
\label{eq:GeneralTwoSplitGARK}
\begin{eqnarray}
\label{eqn:imex-components-explicit-stage}
\begin{bmatrix} Y^{\expl}_i \\ Z^{\expl}_i  \end{bmatrix} &=& \begin{bmatrix}  
y_{n} + h\, \sum_{j=1}^{i-1}  a_{i,j}^{\expl,\expl} ~f(T^{\expl}_j, Y^{\expl}_j, Z^{\expl}_j) \\
z_{n} + h\, \sum_{j=1}^{i-1} a_{i,j}^{\expl,\impl} ~g(T^{\impl}_j, Y^{\impl}_j , Z^{\impl}_j ) 
 \end{bmatrix}, \qquad \\
\label{eqn:imex-components-implicit-stage}
\begin{bmatrix} Y^{\impl}_i \\  Z^{\impl}_i  \end{bmatrix}  &=& \begin{bmatrix}  
y_{n} + h\,  \sum_{j=1}^{i} a_{i,j}^{\impl,\expl} ~ f(T^{\expl}_j, Y^{\expl}_j, Z^{\expl}_j) \\
z_{n} + h\,  \sum_{j=1}^{i}  a_{i,j}^{\impl,\impl}~ g(T^{\impl}_j, Y^{\impl}_j , Z^{\impl}_j)
\end{bmatrix},\\
\label{eqn:imex-components-sol}
\begin{bmatrix}   y_{n+1} \\ z_{n+1}  \end{bmatrix}  &=& \begin{bmatrix}  
y_{n} + h\,  \sum_{i=1}^{s^{\expl}}  b_{i}^{\expl}~ f(T^{\expl}_i, Y^{\expl}_i, Z^{\expl}_i) \\ 
z_{n}  + h\,  \sum_{i=1}^{s^{\impl}} b_{i}^{\impl} ~ g(T^{\impl}_i, Y^{\impl}_i , Z^{\impl}_i)
\end{bmatrix}.  
\end{eqnarray}
Here $(\mathbf{A}^{\expl,\expl} ,\mathbf{b}^{\expl},\mathbf{c}^{\expl})$ is an explicit Runge-Kutta method ($a^{\expl,\expl}_{i,j} = 0$ for $j \geq i$) used to solve the non-stiff component.
Similarly, $(\mathbf{A}^{\impl,\impl} ,\mathbf{b}^{\impl},\mathbf{c}^{\impl})$ is a diagonally implicit Runge-Kutta method ($a^{\impl,\impl}_{i,j} = 0$ for $j > i$) used to solve the stiff component.
The matrices $\mathbf{A}^{\expl,\impl}$ and $\mathbf{A}^{\impl,\expl}$ realize the coupling between the non-stiff and the stiff components.  The time argument is evaluated as 
\begin{equation}
\label{eq:Tnj}
    T^{\expl}_j = t_n +h {c}^{\expl}_j, ~  T^{\impl}_j = t_n +h {c}^{\impl}_j.
\end{equation}
This method can be summarized in the following Butcher tableau: 
\begin{equation}
\label{eq:GARKgeneral}
    \begin{array}{c|c|c}
        \mathbf{c}^{\expl} & \mathbf{A}^{\expl,\expl} & \mathbf{A}^{\expl,\impl}  \\
        \hline
        \mathbf{c}^{\impl} & \mathbf{A}^{\impl,\expl} & \mathbf{A}^{\impl,\impl}  \\
        \hline
         & \mathbf{b}^{\expl\,T} & \mathbf{b}^{\impl\,T}   
    \end{array},
\end{equation}
\end{subequations}
with the coefficient matrices $\mathbf{A}^{u,v} \in \mathbb{R}^{s^{u} \times s^{v}}$, $\mathbf{b}^{u} \in \mathbb{R}^{s^u}$, $\mathbf{c}^{u} =  \mathbf{A}^{u,v} \cdot \mathbf{1}^{v} \in \mathbb{R}^{s^u}$, and $\mathbf{1}^{v} \in \mathbb{R}^{s^v}$ a vector of ones, for $u,v \in \{\expl,\impl\}$. The internal consistency conditions $\mathbf{A}^{\expl,\impl}\cdot \mathbf{1}^{\impl} = \mathbf{c}^{\expl}$ and $\mathbf{A}^{\impl,\expl}\cdot \mathbf{1}^{\expl} = \mathbf{c}^{\impl}$ are assumed to be satisfied (\cite{GARKSandu}).

As an example, the IMEX Euler method in \cref{eq:IMEXExplicit first} is a GARK scheme of the form \cref{eq:GeneralTwoSplitGARK} characterized by the following Butcher tableau:
\begin{equation}\label{eq:AdamGARKOriginal}
\garkmethod{0}%
{0}%
{0}%
{1}%
{1}%
{1}%
{1}%
{1}.
\end{equation}
%
%
%
\subsection{\textsc{Adam} as  the IMEX Euler Discretization} 
\label{subsec:Adam_IMEX_Euler}
We will now demonstrate how the \textsc{Adam} method fits in the GARK scheme \cref{eq:GARKgeneral}.
We perform a component splitting of the \textsc{Adam} ODE \cref{eq:ODE} as follows:
\[
z =  \begin{bmatrix} \theta \\ t \end{bmatrix}, \quad y = \begin{bmatrix} m \\ v \end{bmatrix},
\]
and write the system  \cref{eq:ODE}  in partitioned form \cref{eq:ComponentSplit_IMEX}:
%
\begin{equation}
\label{eq:Adam-partitioned-ODE}
\begin{bmatrix} \dot{y} \\ \dot{z} \end{bmatrix} =
    \begin{bmatrix} \dot{m} \\ \dot{v}  \\  \dot{\theta} \\ \dot{t}  \end{bmatrix} = 
    \begin{bmatrix} 
    d(t) ~\nabla_\theta\mathcal{L} (t,\theta)  - r(t)~ m(t) \\ 
    p(t) ~ [\nabla_\theta \mathcal{L}(t,\theta)]^{\circ 2} - q(t)~ v(t) \\
    - m(t)/\sqrt[\circ]{v(t) + \epsilon} \\
    1 \end{bmatrix}
    =
    \begin{bmatrix} f(y,z) \\ g(y,z) \end{bmatrix},
\end{equation}
where we used notation adopted in  \cref{eqn:define-L}.
Solving the partitioned system \cref{eq:Adam-partitioned-ODE} with the IMEX Euler method \cref{eq:IMEXExplicit first} with a step size $h$ gives:
\begin{equation}
\label{eq:EulerAdam}
\begin{split}
    \begin{bmatrix}  m_{n+1} \\ v_{n+1}  \\ \theta_{n+1} \\ t_{n+1}  \end{bmatrix} = 
    \begin{bmatrix} 
    m_{n} + h\, d(t_{n+1}) ~\nabla_\theta\mathcal{L}(t_{n+1},\theta_{n})  - h\,r(t_n)~ m_{n} \\ 
    v_{n} + h\,p(t_{n+1}) ~ [\nabla_\theta \mathcal{L}(t_{n+1},\theta_{n})]^{\circ 2} - h\,q(t_n)~ v_{n} \\ 
    \theta_{n} - h\,m_{n+1}/\sqrt[\circ]{v_{n+1} + \epsilon} \\
    t_{n} + h\,
    \end{bmatrix}. 
    \end{split}
\end{equation}
Note that although this is an application of the IMEX Euler method, no nonlinear systems of equations are solved in \cref{eq:EulerAdam}.
This is due to the special structure of \cref{eq:Adam-partitioned-ODE}, where the stiff dynamics $g(y,z)$ does not depend on $\theta$, therefore $\theta_{n+1}$ is not present on the right hand side of the third equation in \cref{eq:EulerAdam}, and can be computed via an explicit update.

We compare \cref{eq:EulerAdam} with the standard \textsc{Adam}.
Identifying the \textsc{Adam} parameters as $\beta_1 = 1 -  h\, r_{n} = 1 - h\, d_{n}$, $\beta_2 =  1 - h\, q_{n} = 1 - h\, p_{n}$, 
we conclude that the IMEX Euler solution \cref{eq:EulerAdam} of the optimization ODE \cref{eq:Adam-partitioned-ODE} recovers the original discrete \textsc{Adam} method. 
In other words, \textsc{Adam} is the IMEX Euler discretization of \cref{eq:Adam-partitioned-ODE}.
\subsection{Linear Stability of the IMEX Euler discretization of the optimization ODE}\label{subsec:linear_stability_imex_euler}

Similar to \cref{subsec:linear_stability_FE}, we can perform a  linearization of the \cref{eq:EulerAdam} method. Considering  $\nabla_\theta\mathcal{L}(t_{n},\theta_{n})] \approx \Lambda  \theta_n$, $p_n = p, q_n =q , d_n =d, r_n=r$ for all timesteps and dropping all nonlinear terms reveals the following stability matrix for the IMEX Euler scheme applied to the optimization ODE:
\begin{equation}
    \label{eq:IMEXEulerStability}
    \begin{bmatrix}
        \theta \\ m \\ v
    \end{bmatrix}_{n+1} = 
    \left[\begin{array}{ccc}1-\frac{dh^2L}{\sqrt{\epsilon}}&-\frac{h(1-hr)}{\sqrt{\epsilon}}&0\\dhL&1-hr&0\\0&0&1-hq\\\end{array}\right]
        \begin{bmatrix}
        \theta \\ m \\ v
    \end{bmatrix}_{n}.
\end{equation}
The eigenvalues of the stability matrix are:
\begin{equation}
    \left\{1-hq, \frac{-d h^2 L \sqrt{\epsilon }-h r
   \epsilon +2 \epsilon  \pm \sqrt{\left(d h^2 L \sqrt{\epsilon }+h r \epsilon -2 \epsilon
   \right)^2-4 \left(\epsilon ^2-h r \epsilon ^2\right)}}{2 \epsilon }\right\}.
\end{equation} We note that the IMEX Euler method is still conditionally stable but with a different stability criterion. For a stability region plot of this method under nominal hyperparameters, refer to \cref{app:stability_plots}.
\subsection{Improving \textsc{Adam} via High Order IMEX GARK Discretizations} 
\label{Adam_IMEX_GARK}

When initial value problems are considered, higher-order time integration schemes outperform the first-order Euler method.
Higher-order methods have smaller local and global truncation errors and require fewer steps to reach a desired level of accuracy compared to lower-order methods.
We make the ansatz that improved versions of the \textsc{Adam} can be obtained by applying higher order IMEX GARK schemes to the partitioned optimization ODE \cref{eq:Adam-partitioned-ODE}.
Specifically, we investigate how higher-order IMEX time discretization methods lead to better stochastic optimization algorithms, i.e., result in a faster decrease of the loss function toward a local minimum.

Application of an IMEX GARK scheme \cref{eq:GeneralTwoSplitGARK} to solve the partitioned \textsc{Adam} ODE system \cref{eq:Adam-partitioned-ODE} leads to the following discrete optimization algorithm \cite{GARKSandu} :
\begin{equation}
\label{eq:GARKAdamweightsmomentumvelocity}
    \begin{split}
        m^{\expl}_{i} &= m_{n} + h \sum_{j = 1}^{i-1} a^{\expl,\expl}_{i,j}\left( d(T^{\expl}_j) \nabla_\theta\mathcal{L}(T^{\expl}_j, \theta^{\expl}_j) - r(T^{\expl}_j) m^{\expl}_j\right),\quad i = 1, \dots, s^{\expl},\\
        v^{\expl}_{i} &= v_{n} + h \sum_{j = 1}^{i-1} a^{\expl,\expl}_{i,j}\left( p(T^{\expl}_j) \nabla_\theta\mathcal{L} (T^{\expl}_j, \theta^{\expl}_j)^{\circ 2} - q(T^{\expl}_j) v^{\expl}_{j}\right),\quad  i = 1, \dots, s^{\expl},\\
        m^{\impl}_{i} &= m_{n} + h \sum_{j = 1}^{i} a^{\impl,\expl}_{i,j}\left( d(T^{\expl}_j)  \nabla_\theta\mathcal{L}(T^{\expl}_j,\theta^{\expl}_{j}) - r(T^{\expl}_j) m^{\expl}_j\right),\quad  i = 1, \dots, s^{\impl},\\
         v^{\impl}_{i} &= v_{n} + h \sum_{j = 1}^{i} a^{\impl,\expl}_{i,j}\left(p(T^{\expl}_j) \nabla_\theta\mathcal{L}(T^{\expl}_j,\theta^{\expl}_{j})^{\circ 2} - q(T^{\expl}_j) v^{\expl}_{j}\right),\quad  i = 1, \dots, s^{\impl},\\
         \theta^{\expl}_{i} &= \theta_{n} + h \sum_{j = 1}^{i-1} a^{\expl,\impl}_{i,j} \frac{-m^{\impl}_{j}}{\sqrt[\circ]{v^{\impl}_{j} + \varepsilon}},\quad i = 1, \dots, s^{\expl},\\
        m_{n+1} &= m_{n} + h\sum_{i=1}^{s^{\expl}}b_{i}^{\expl} \left(d(T^{\expl}_j) \nabla_\theta\mathcal{L}(T^{\expl}_i,\theta^{\expl}_{i}) - r(T^{\expl}_i) m^{\expl}_{i}\right),\\
         v_{n+1} &= v_{n} + h\sum_{i=1}^{s^{\expl}}b_{i}^{\expl} \left(p(T^{\expl}_i) \nabla_\theta\mathcal{L}(T^{\expl}_i,\theta^{\expl}_{i})^{\circ 2} - q(T^{\expl}_i) v^{\expl}_{i}\right), \\
          \theta_{n+1} &= \theta_{n} + h\sum_{i=1}^{s^{\impl}}b_i^{\impl}  \frac{-m^{\impl}_{i}}{\sqrt[\circ]{v^{\impl}_{i} + \varepsilon}}, \\
          t_{n+1} &= t_{n} + h.
    \end{split}
\end{equation}
where $T_j^{\expl}$ and $T_j^{\impl}$ are defined in \cref{eq:Tnj}.

We reiterate the fact that formulation in \cref{eq:GARKAdamweightsmomentumvelocity} does not require any implicit (algebraic) solve since the update to $\theta$ in \cref{eq:EulerAdam} does not depend on $\theta$.
This is important, as we do not want to incur the extra cost of solving a non-linear equation during training on large datasets. 

The coefficients of an IMEX GARK scheme employed in \cref{eq:GARKAdamweightsmomentumvelocity} need to satisfy the order conditions developed in \citep{GARKSandu}.
The IMEX Euler coefficients \cref{eq:AdamGARKOriginal} applied to \cref{eq:GARKAdamweightsmomentumvelocity} form a first order IMEX GARK, and recover the original \textsc{Adam} method \citep{kingma2017Adam}.

In this work, we focus on a second-order IMEX GARK method that employs the implicit trapezoidal scheme (often called Crank Nicholson) for the stiff component and the explicit trapezoidal scheme for the non-stiff variables.
This method, called ``IMEX Trapezoidal'' is defined by the following Butcher tableau of coefficients as developed in \cref{eq:GARKgeneral}: 
\begin{equation}
\label{eq:AdamGARKTrapezoidal}
    \begin{array}{c|c|c}
        \mathbf{c}^{\expl} & \mathbf{A}^{\expl,\expl} & \mathbf{A}^{\expl,\impl}  \\
        \hline
        \mathbf{c}^{\impl} & \mathbf{A}^{\impl,\expl} & \mathbf{A}^{\impl,\impl}  \\
        \hline
         & \mathbf{b}^{\expl\,T} & \mathbf{b}^{\impl\,T}   
    \end{array}
=
\garkmethod{0 \\ 1}%
{ 0 & 0 \\ 1 & 0 }%
{ 0 & 0  \\1 & 0 }%
{0 \\ 1}%
{0 & 0 \\ \frac{1}{2} & \frac{1}{2}}%
{0 & 0\\ \frac{1}{2} & \frac{1}{2} }%
{\frac{1}{2} & \frac{1}{2}}%
{\frac{1}{2} & \frac{1}{2}}.
\end{equation}
Application of the IMEX Trapezoidal method \cref{eq:AdamGARKTrapezoidal} applied to solve the
\textsc{Adam} partitioned ODE \cref{eq:Adam-partitioned-ODE} leads to the new ``IMEX Trapezoidal \textsc{Adam}'' optimization scheme.
The computational steps of this method are given explicitly in \Cref{alg:IMEXTRAP}.
This optimization scheme performs well on several classification and regression problems, as confirmed by numerical experiments reported in \Cref{sec:numerical_experiments}.
The second-order IMEX method achieves a good balance between performance and the number of gradient evaluations for several test cases.

\begin{algorithm}[H]
    \SetAlgoLined
    \KwData{Stepsize $h$; $\beta_1, \beta_2 \in [0,1)$, Loss function $\mathcal{L}(t,\theta)$ }
    \KwResult{Optimal parameters $\theta^*$}
    initialize $n \gets 0, m_0 \gets 0, v_0 \gets g_{0}^{\circ 2} ~ \text{where}~ g_0 = \nabla \mathcal{L}(t_0, \theta_0)$\;
    \While{$\theta_n$ not converged}{
        $k_{1}^{m} \gets \frac{1}{\alpha_1} [\nabla_{\theta} \mathcal{L}(t_{n+1}; \theta_{n}) - m_{n}]$\;
        $k_{1}^{v} \gets \frac{1}{\alpha_2} [\nabla_{\theta} \mathcal{L}^{\circ 2}(t_{n+1}; \theta_{n}) - v_{n}]$\; 
        $k_{1}^{\theta} \gets - \dfrac{m_{n}}{\sqrt{v_{n} + \epsilon}}$\;
        $k_{2}^{m} \gets \frac{1}{\alpha_1} [\nabla_{\theta} \mathcal{L}(t_{n+1}; \theta_{n} + h\, k_{1}^{\theta}) - (m_{n} + h~ k_{1}^{m})]$\;
        $k_{2}^{v} \gets \frac{1}{\alpha_2} [\nabla_{\theta} \mathcal{L}^{\circ 2}(t_{n+1}; \theta_{n} + ~ h \, k_{1}^{\theta}) - (v_{n} + ~ h \, k_{1}^{v})]$\; 
        $k_{2}^{\theta} \gets - \dfrac{m_{n} + 0.5 ~ h\, k_{1}^{m} + 0.5 ~ h\, k_{2}^{m}}{\sqrt[\circ]{v_{n} + 0.5 ~ h\, k_{1}^{v} + 0.5 ~ h\, k_{2}^{v} + \epsilon}}$\;
        $\theta_{n+1} \gets \theta_n + 0.5 ~h\, k_{1}^{\theta} + 0.5 ~h\, k_{2}^{\theta}$\;
        $m_{n+1} \gets m_n + 0.5 ~h\, k_{1}^{m} + 0.5 ~h\, k_{2}^{m}$\;
        $v_{n+1} \gets v_n + 0.5 ~h\, k_{1}^{v} + 0.5 ~h\, k_{2}^{v}$\;
        $n \gets n+1$.
    }
\caption{IMEX Trapezoidal \textsc{Adam} Optimizer}\label{alg:IMEXTRAP}
\end{algorithm}

\section{Numerical Experiments}
\label{sec:numerical_experiments}











This section provides a number of experiments that involve regression and classification tasks, chosen based on their relevance in machine learning and scientific computing applications.
A grid search is performed to tune the hyperparameters in the following experiments. The effects of initialization of learnable parameters on the convergence to the local minima are well known in ML literature \citep{initialization_and_momentum_in_deep_learning, InitializationSummary}. 
We resort to Xavier Initialization \citep{glorotinitialization} in this study for all the experiments .
To ensure robust results regardless of random initialization, every experiment is randomly initialized $20$ times, and the average loss value is reported. 

The momentum is initialized to zero and the velocity to the square of the initial gradient for all test problems.
Our experiments focus on determining the time-stepping method's ability to minimize the cost function during training and not the generalization error on test data. Nevertheless, we report test losses in several cases.
Methods such as early stopping, drop out \cite{dropout}, and regularization can be used to mitigate over-fitting and do not conflict with our proposed method.
The Mean Square Error (MSE) loss function is used for regression, and Categorical Cross-Entropy loss is used for classification problems.
The decrease in the cost function is measured against the total number of gradient evaluations to measure the optimizer's effectiveness compared to the computational costs incurred.
Since higher-order methods have more stages, they use more gradient evaluations in each epoch (or time step). Therefore, comparing methods based on the total gradient evaluations is a fair assessment criterion.

Stochastic Gradient Descent (\textsc{SGD}), Forward Euler (FE ), IMEX \textsc{Adam}, and IMEX Trapezoidal methods are compared for each test case.
We also tested a number of higher-order Runge Kutta (RK), and IMEX GARK methods. However, they did not consistently perform well in all experiments, and further study is needed to determine whether they pose any real advantage for the increased number of gradient evaluations they incur. Interested readers can refer to \ref{app:higher_order_methods} for these results.


\subsection{Lorenz '63 Dynamics}
\label{subsec:L63_Exp}


The Lorenz '63 model \cite{Lorenz63} is a nonlinear chaotic system represented by equations:
\begin{equation}\label{eq:L63}
    \begin{split}
        \dot{x} &= \sigma (y-x), \\ 
        \dot{y} &= x(\rho - z) - y, \\
        \dot{z} &= xy - \beta z,
    \end{split}
\end{equation}
and parameters $\sigma = 10, \rho = 8/3, \beta = 28$. Due to its chaotic nature, it is consistently used in scientific machine learning to benchmark different methodologies, e.g., see \cite{brunton2016discovering,de2023ai,fablet2021learning}.

Consolidating all the states $[x,y,z]_{n}$ into input $S_n$, the neural network approximates the state $S_{n+1}$ at the subsequent time-step.
The architecture has a single hidden layer with 100 neurons, a constant learning rate of $0.01$, a hyperbolic tangent activation function ($\tanh$), and uses parameters $\beta_1 = 0.9$ and $\beta_2 = 0.95$.

\begin{figure}[h]
\centering
\begin{subfigure}[b]{0.44\textwidth}
\includegraphics[width=\textwidth]{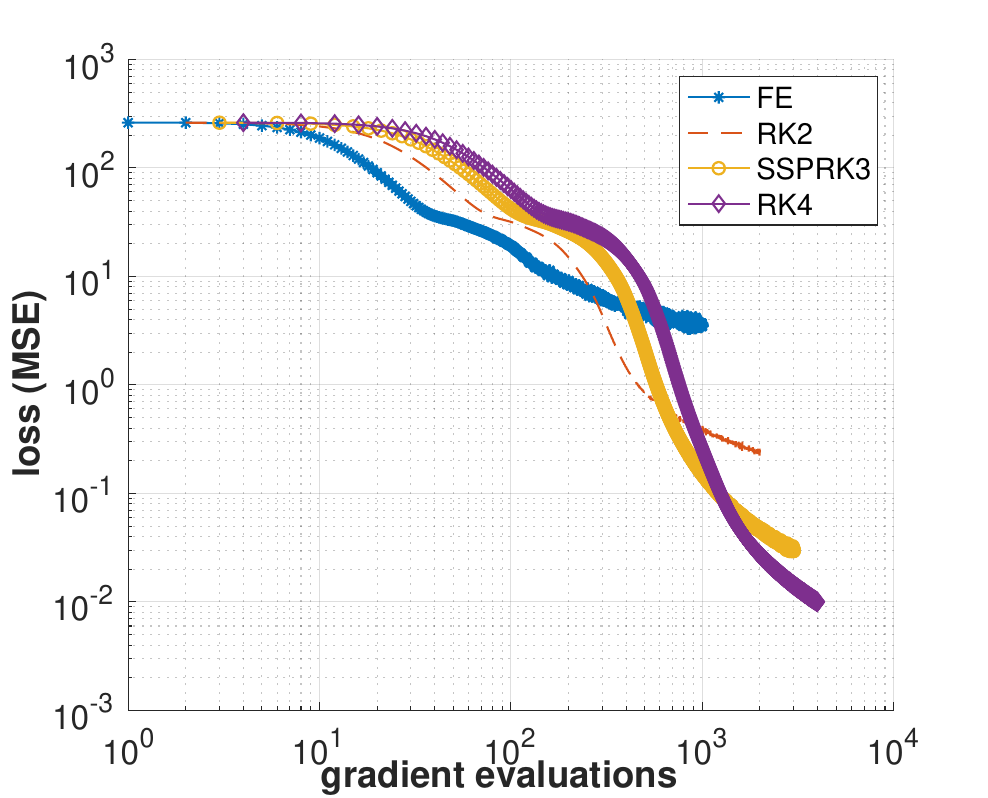}
\caption{batch = $1$}
\label{fig:Fig1_Supp_L63}
\end{subfigure}
\begin{subfigure}[b]{0.44\textwidth}
\includegraphics[width=\textwidth]{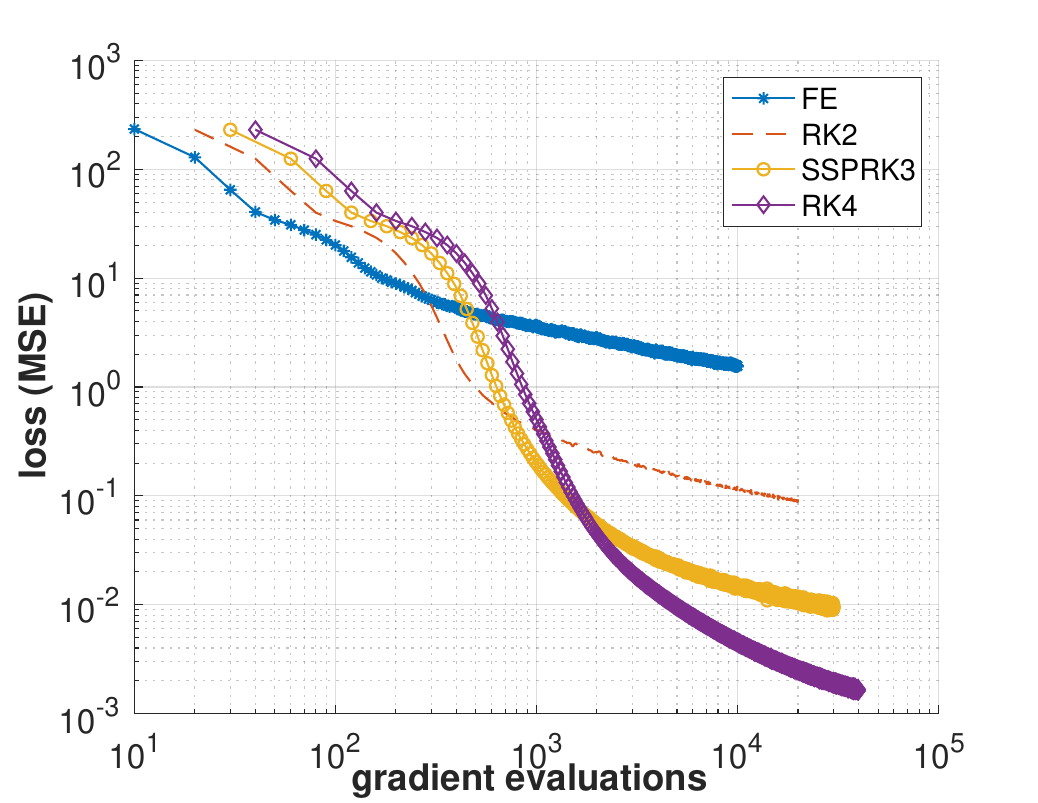}
\caption{batch = $10$}
\label{fig:Fig2_Supp_L63}
\end{subfigure}
\\
\begin{subfigure}[b]{0.44\textwidth}
\includegraphics[width=\textwidth]{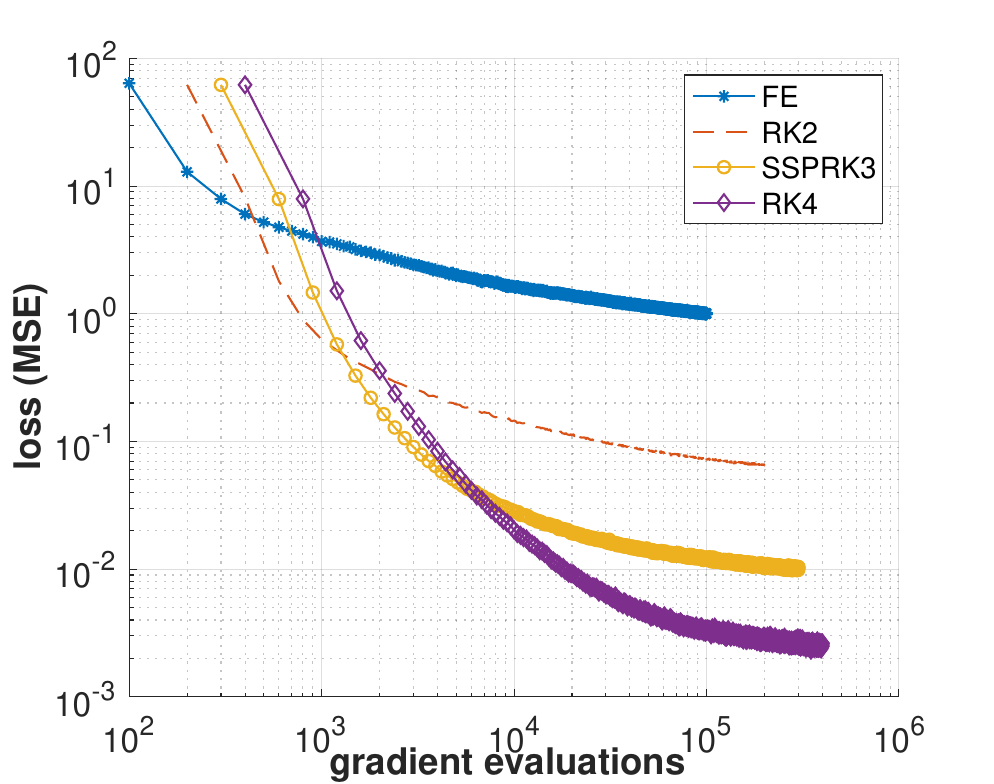}
\caption{batch = $100$}
\label{fig:Fig3_Supp_L63}
\end{subfigure}
\begin{subfigure}[b]{0.44\textwidth}
\includegraphics[width=\textwidth]{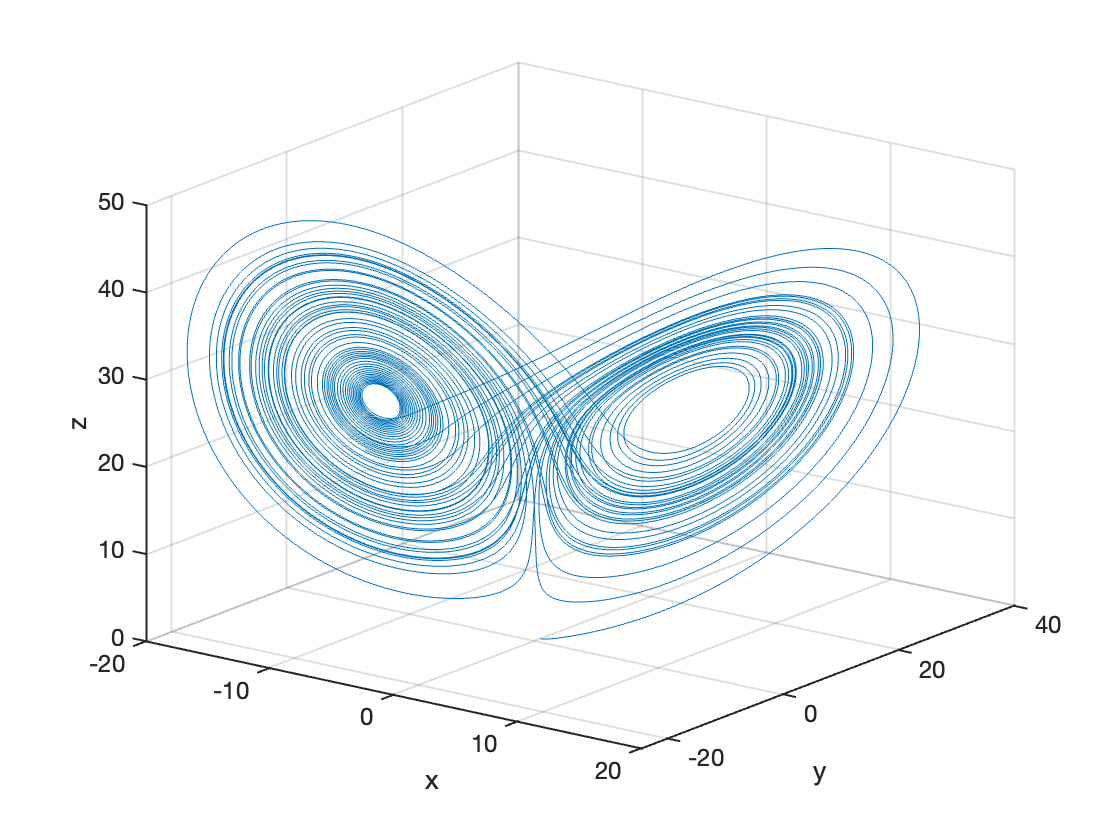}
\caption{Lorenz Butterfly}
\label{fig:Fig0_Supp_L63}
\end{subfigure}
\caption{Regression on learning the L63 chaotic dynamics. The learning rate = $0.01$, epochs = $1000$, $\beta_1 = 0.9$, $\beta_2 = 0.95$ in the training process. Methods of orders one (FE), two (RK2), three(SSPRK3), and four (RK4) are compared. RK4 outperforms all the other methods for a comparable number of gradient evaluations for both batched and mini-batched training. 
}  
\label{fig:Supp_L63}
\end{figure}

\begin{figure}[ht]
\vspace{.3in}
\centering
\includegraphics[width=0.5\textwidth]{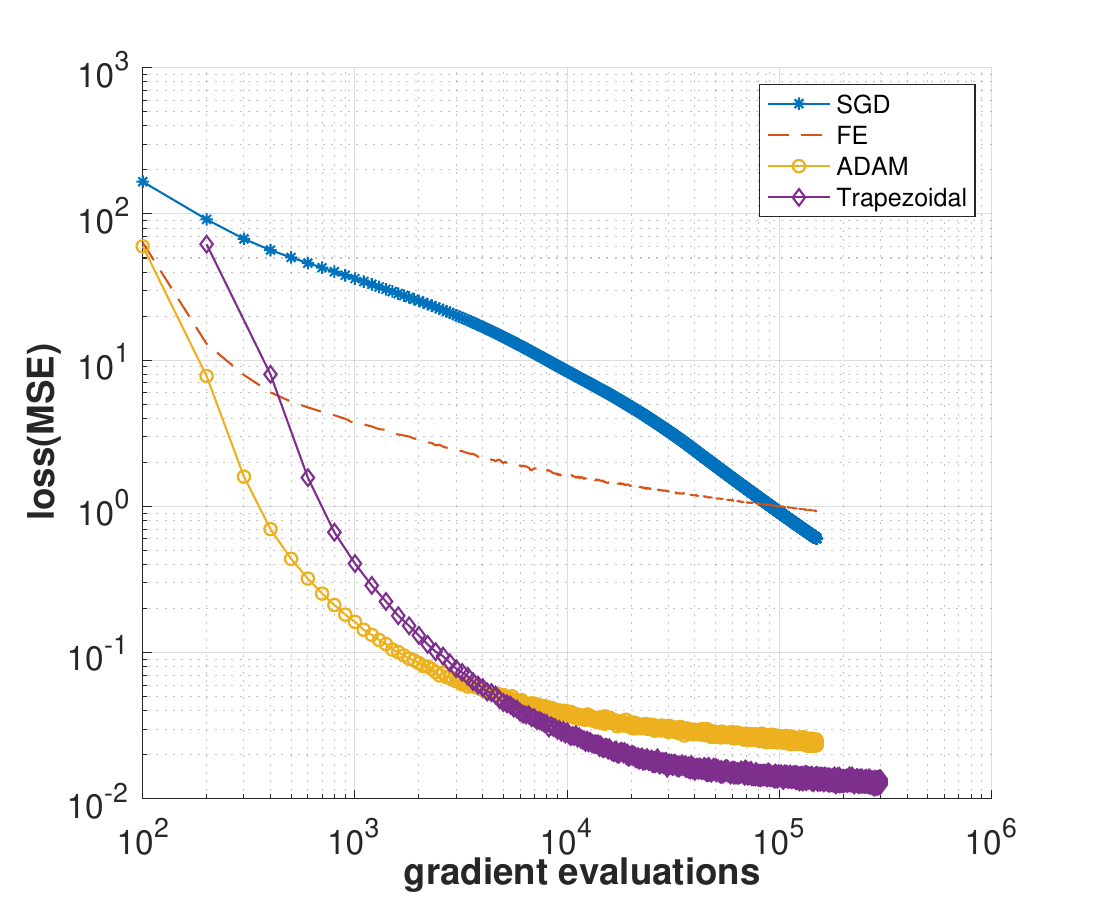}
\vspace{.3in}
\caption{Regression on learning the L63 chaotic dynamics. The learning rate = $0.01$, batches = $100$, epochs = $1500$, $\beta_1 = 0.9$, $\beta_2 = 0.95$ in the training process. \textsc{SGD}, Forward Euler, IMEX \textsc{Adam} and IMEX Trapezoidal discretization of \cref{eq:ODE} are compared. IMEX Trapezoidal seems to outperform the original \textsc{Adam} as learning proceeds for comparable number of gradient evaluation.}
\label{fig:F1_L63}
\end{figure}

The complete dataset consists of $10,000$ points created using \texttt{ODE Test Problems} package by \cite{otp} written in Matlab.
The first set of results in \Cref{fig:Supp_L63} compare the traditional time stepping methods such as Forward Euler, Heun's Method, Stability Preserving Third Order Runge Kutta (SSPRK3) and Fourth Order Runge Kutta (RK4) applied to this regression problem.
The $x-$axis in the plots represents the total number of gradient evaluations, and the $y-$axis denotes the value of the cost function. 
\Cref{fig:Fig1_Supp_L63} shows a non-stochastic training process without batching. 
\Cref{fig:Fig2_Supp_L63} and \cref{fig:Fig3_Supp_L63} denotes training in batches of data while \cref{fig:Fig0_Supp_L63} shows the well-known ``butterfly'' shape of the solution trajectory.
These plots show the preliminary indication that higher-order methods yield faster convergence to the minimum.
We continue the experiments in  \Cref{fig:F1_L63}, in which we compare the new IMEX method against standard optimization methods. \Cref{fig:F1_L63} shows the fast convergence of the IMEX methods compared to \textsc{SGD} and \textsc{FE}. The trapezoidal method converges to a smaller loss function, beating  \textsc{Adam} towards the end of the training. 

\subsection{Sum of Gaussians}\label{subsec:G3_Exp}

\begin{figure}[h]
\centering
\vspace{.3in}
\begin{subfigure}[b]{0.32\textwidth}
\includegraphics[width=\textwidth]{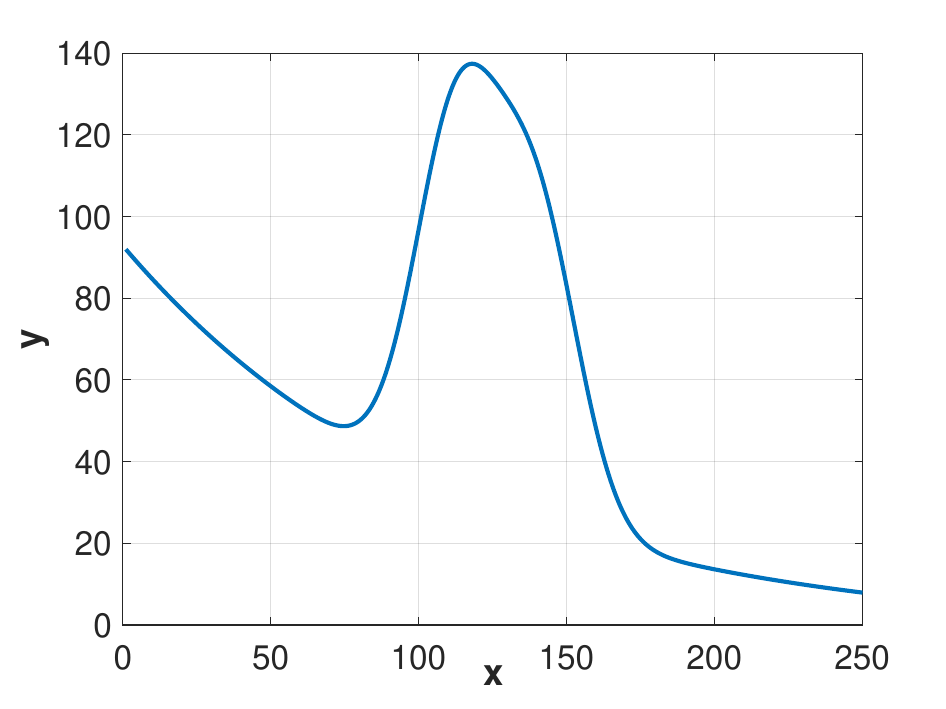}
\caption{Gauss3 contour}
\label{fig:Fig1_Supp_G3}
\end{subfigure}
\begin{subfigure}[b]{0.32\textwidth}
\includegraphics[width=\textwidth]{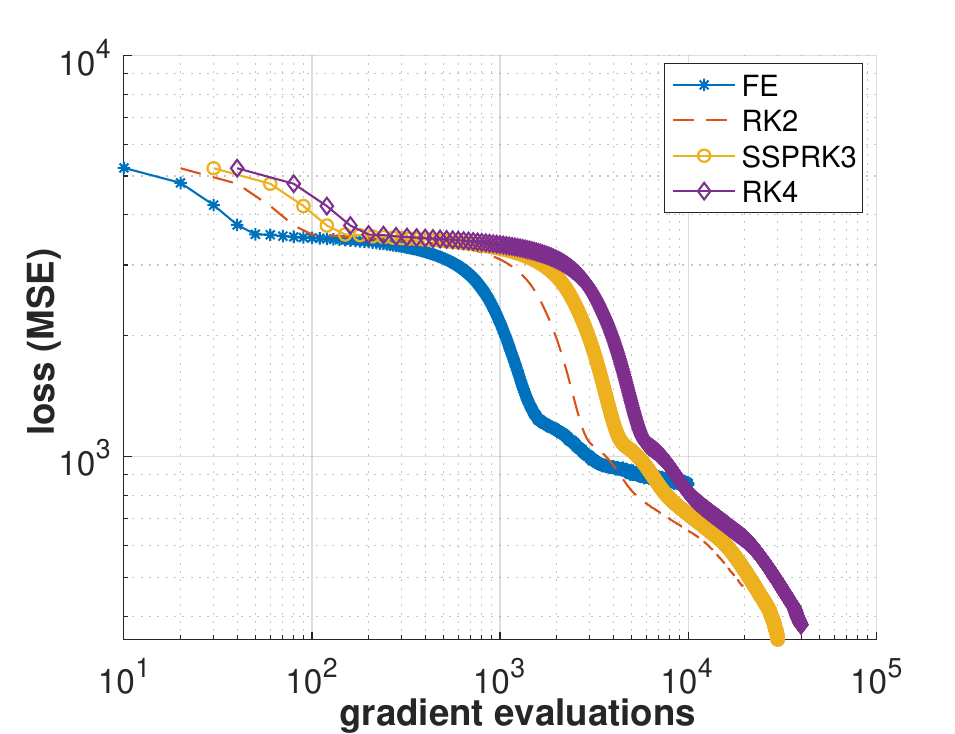}
\caption{batch = $10$}
\label{fig:Fig2_Supp_G3}
\end{subfigure}
\begin{subfigure}[b]{0.32\textwidth}
\includegraphics[width=\textwidth]{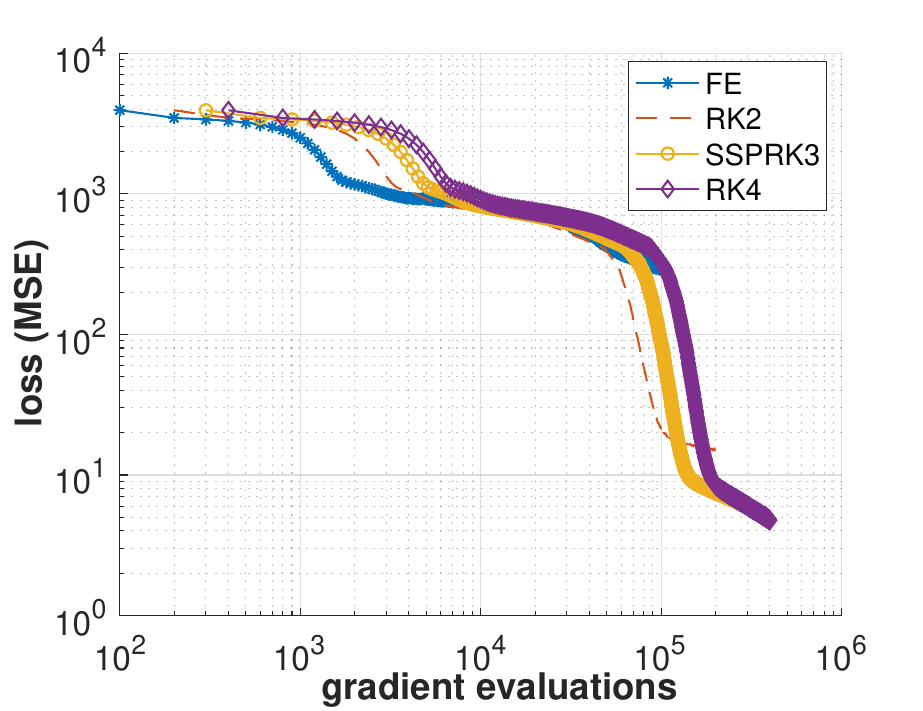}
\caption{batch = $100$}
\label{fig:Fig3_Supp_G3}
\end{subfigure}
\vspace{.3in}
\caption{Nonlinear Regression on the NIST Gauss3 dataset with a shallow network and varying stochastic noise. Learning rate = $0.001$, $\beta_1 = 0.9$, $\beta_2 = 0.9$, epochs = $1000$ with $1$ hidden layer ($100$ neurons). \Cref{fig:Fig1_Supp_G3} shows the solution in the range $0 \leq x \leq 250$, \cref{fig:Fig2_Supp_G3} denotes the learning curve with $10$ batches of data while \cref{fig:Fig3_Supp_G3} represents a more stochastic training process. The higher RK methods seem to have an advantage over lower-order ones for this problem setup.}
\label{fig:Supp_G3}
\end{figure}

\begin{figure}[h]
\centering
\begin{subfigure}[b]{0.32\textwidth}
\includegraphics[width=\textwidth]{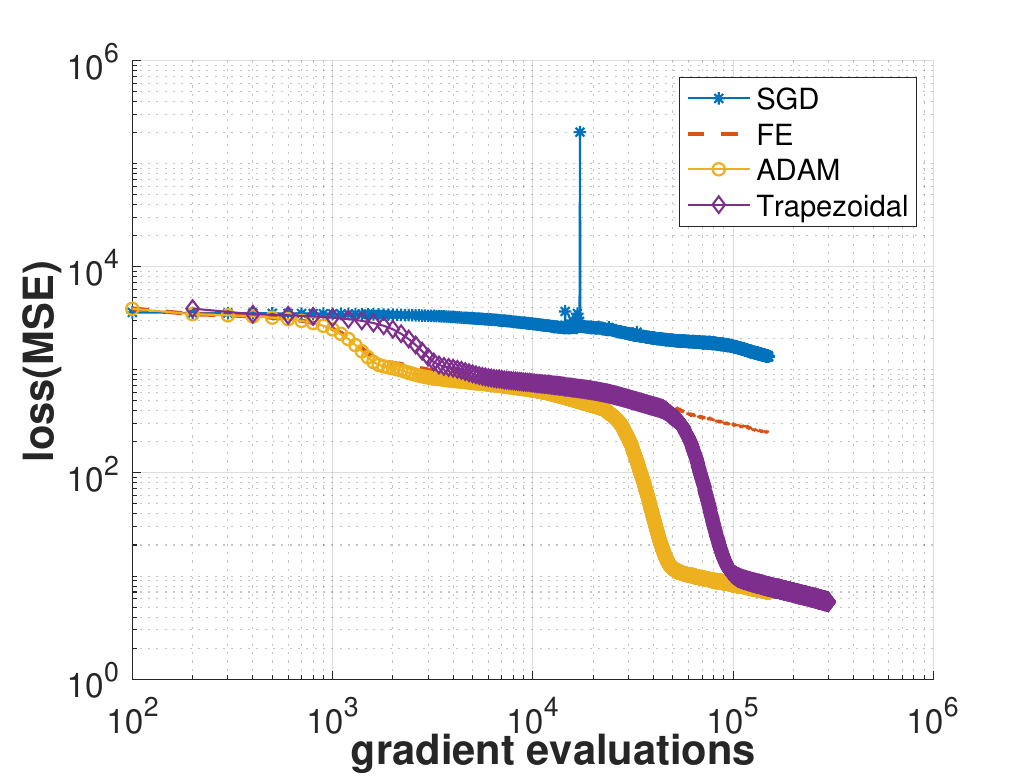}
\caption{Shallow Network,\\ lr = $0.001$}
\label{fig:Fig1_G3}
\end{subfigure}
\begin{subfigure}[b]{0.32\textwidth}
\includegraphics[width=\textwidth]{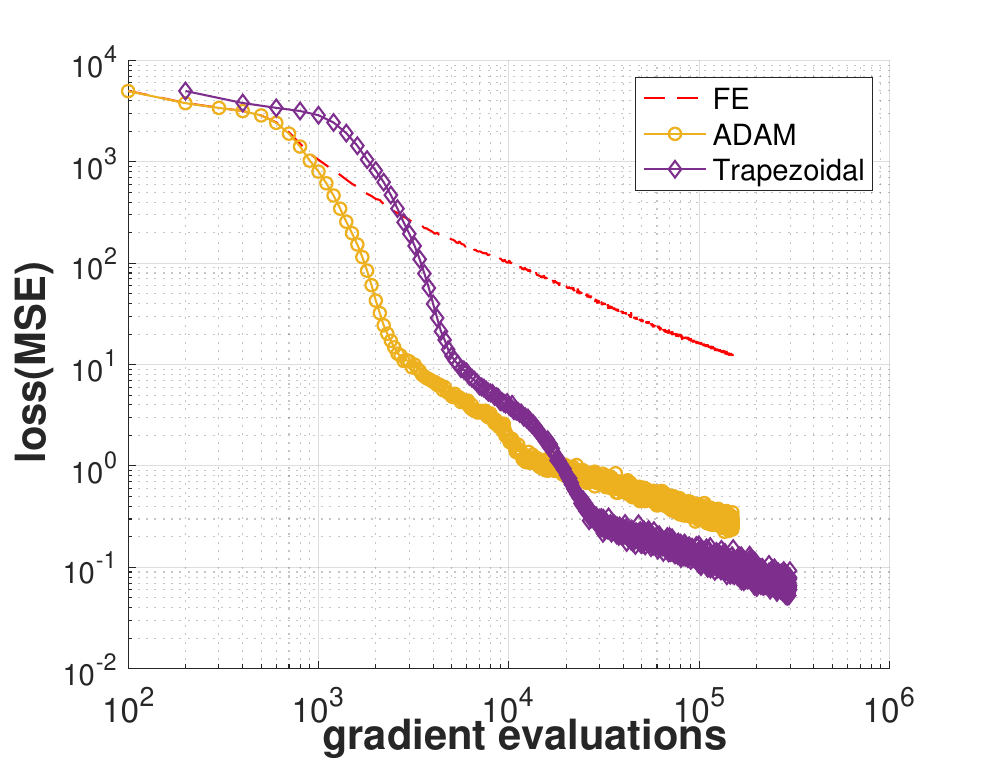}
\caption{Deep Network,\\ lr = $0.001$}
\label{fig:Fig2_G3}
\end{subfigure}
\begin{subfigure}[b]{0.32\textwidth}
\includegraphics[width=\textwidth]{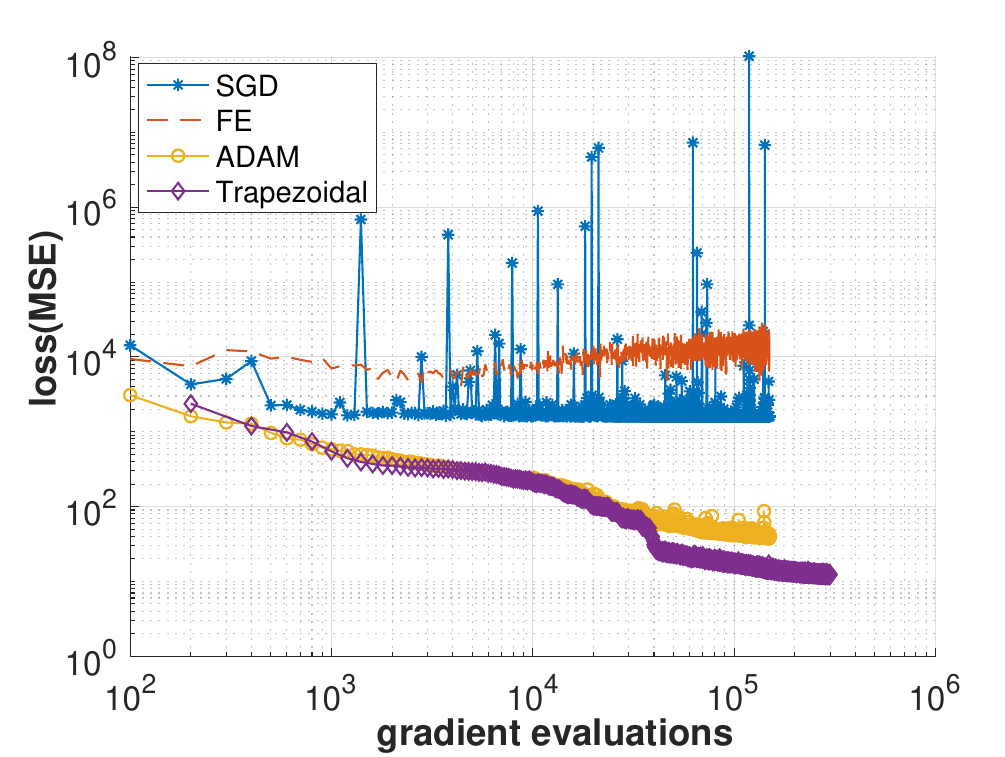}
\caption{Deep Network,\\ lr = $0.1$}
\label{fig:Fig3_G3}
\end{subfigure}
\vspace{.3in}
\caption{Nonlinear regression on the NIST Gauss3 dataset with varying learning rates.  Epochs = 1500, batches = $100$,  $\beta_1 = 0.9$, $\beta_2 = 0.9$, $\epsilon = 10^{-8}$. The loss for a shallow network ($1$ hidden layer with $100$ neurons) with a stable learning rate ($0.001$) is depicted in \cref{fig:Fig1_G3}. A (relatively) deeper network ($5$ hidden layers with $10$ neurons in each) with a stable learning rate = $0.001$ is used in \cref{fig:Fig2_G3}. Finally, some methods become oscilatory with a learning rate = $0.1$ on the deeper network, showcasing the advantage of the higher-order IMEX method over traditional \textsc{SGD} or RK methods.}
\label{fig:Gauss3}
\end{figure}

This relatively difficult nonlinear regression problem is taken from the National Institute of Standards and Technology (NIST) datasets.
We fix the parameters at $b_1 =94.9, b_2=0.009, b_3=90.1, b_4=113, b_5=20, b_6=73.8, b_7=140, b_8=20$ in the target function:
\begin{equation}\label{eq:G3}
\begin{split}
    f(x) = b_1 e^{-b_2 x} + b_3 e^{- \frac{(x-b_4)^2}{(b_5)^2}} + b_6 e^{- \frac{(x-b_7)^2}{(b_8)^2}}, x \in [0, 250].
    \end{split}
\end{equation}
Both shallow and deep feedforward neural networks are used in these experiments.
Initial results with a shallow network are reported in \cref{fig:Supp_G3} and show some promise in using traditional higher-order Runge-Kutta methods. We also note the effects of batch size on the loss curves in \cref{fig:Supp_G3}.
Motivated by these indications, we present the results for the new IMEX methods in \cref{fig:Gauss3}. 

As the loss curves in \cref{fig:Fig3_Supp_G3} denote, the higher-order methods seem to work better in the highly stochastic training case where the batches are smaller and highly randomized. 
 \Cref{fig:Gauss3} is also encouraging as we train on varying depths of neural networks. Fig. \ref{fig:Fig1_G3} shows the result for a shallow network with one layer consisting of 100 neurons. The learning curves show that \textsc{SGD} is not performing well, while both \textsc{Adam} and the Trapezoidal method perform well on this architecture. We note the limited decrease in loss function in the shallow network's case.

The next experiment is with a deep network with 5 hidden layers, each 10 neurons wide, where all other parameters and hyper-parameters are kept the same. We do not show the results for \textsc{SGD} in  \cref{fig:Fig2_G3}, as it fails to reduce the loss in this setup. \textsc{Adam} and the Trapezoidal method perform similarly in the beginning, but the Trapezoidal method performs slightly better as training progresses.
 
Finally, the learning rate for this deep architecture was increased to 0.1. The larger learning rate leads to instability in some methods, as shown in Fig. \cref{fig:Fig3_G3}.
\textsc{SGD} and FE are unstable under such a large learning rate, and the loss values oscillate. Both the IMEX methods are fairly stable.
\subsection{Spiral Dataset}\label{subsec:Spiral_Exp}

\begin{figure}[h]
\centering
\vspace{.3in}
\begin{subfigure}[b]{0.32\textwidth}
\includegraphics[width=\textwidth]{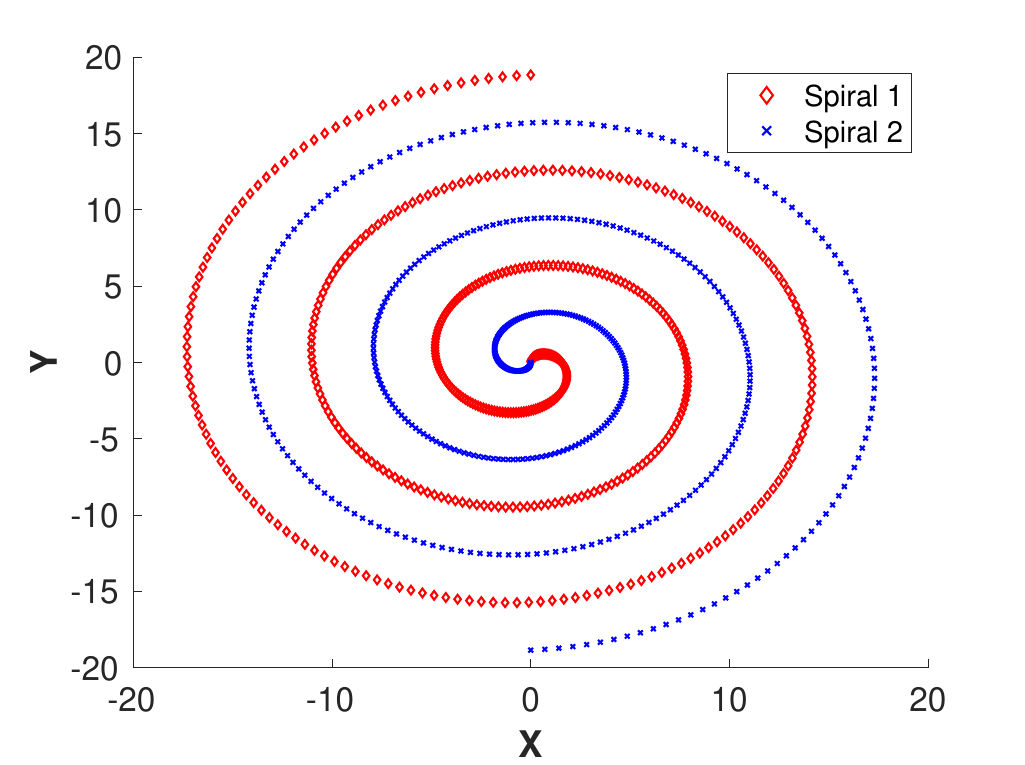}
\caption{Spiral data-set}
\label{fig:Fig1_Supp_Spiral}
\end{subfigure}
\begin{subfigure}[b]{0.32\textwidth}
\includegraphics[width=\textwidth]{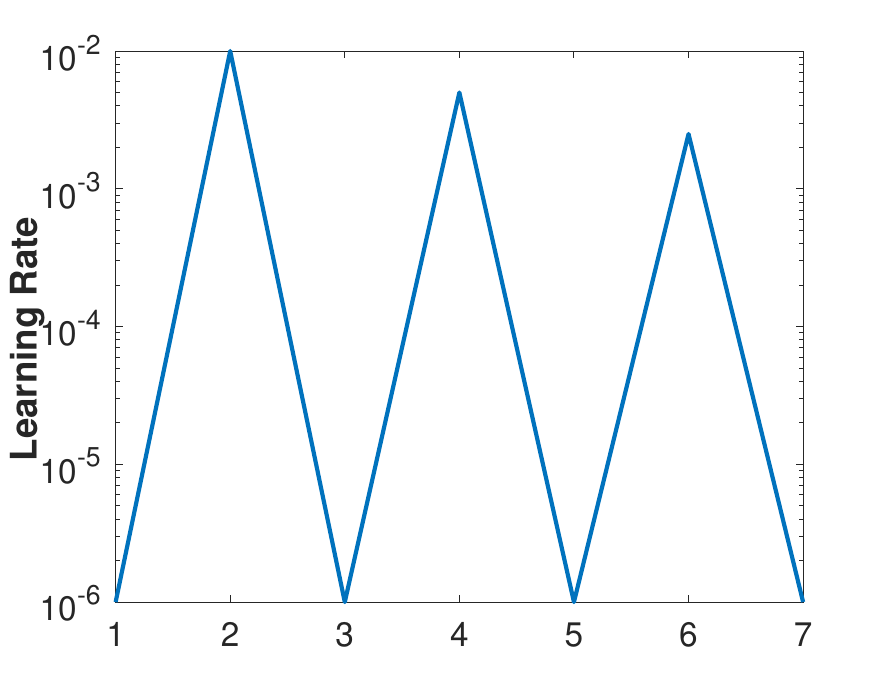}
\caption{Cyclic Learning Rate}
\label{fig:Fig2_Supp_Spiral}
\end{subfigure}
\begin{subfigure}[b]{0.32\textwidth}
\includegraphics[width=\textwidth]{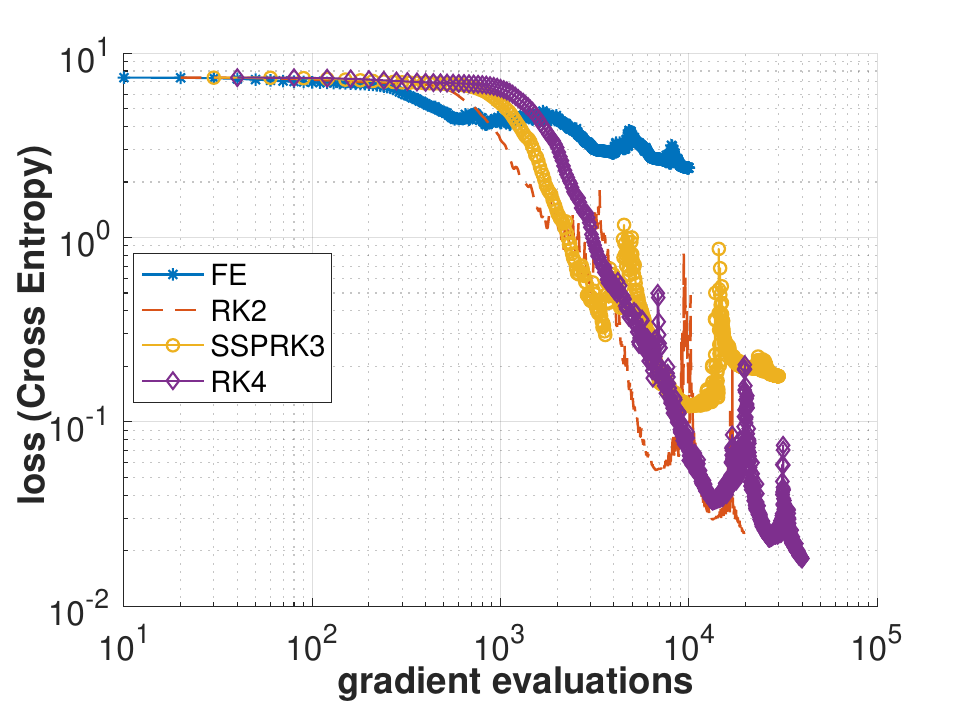}
\caption{Loss}
\label{fig:Fig3_Supp_Spiral}
\end{subfigure}
\vspace{.3in}
\caption{Binary classification on the spiral dataset with a deep network. $\beta_1 = 0.9$, $\beta_2 = 0.999$, batch = $10$ and $8$ hidden layers ($8$ neurons per layer). The dataset and the cyclic learning rate are visualized in \cref{fig:Fig1_Supp_Spiral} and \cref{fig:Fig2_Supp_Spiral} respectively, while the learning curve is shown in \cref{fig:Fig3_Supp_Spiral}. The learning curve shows some advantages of the higher-order RK method compared to the first-order method.}
\label{fig:Supp_Spiral}
\end{figure}

\begin{figure}[h]
\centering
\vspace{.3in}
\includegraphics[width=0.5\textwidth]{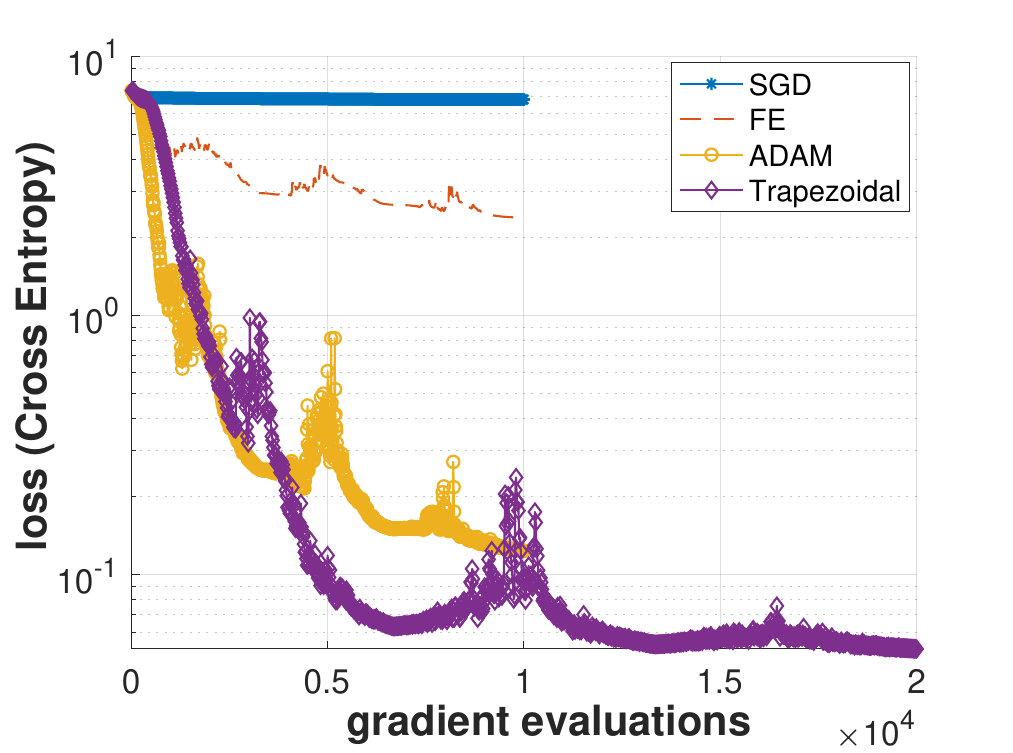}
\caption{Binary classification on the spiral dataset with a deep network ($8$ hidden layers with $8$ neurons in each). Epochs = 1000, batch = 10, $\beta_1 = 0.9, \beta_2 = 0.999, \epsilon = 10^{-8} $.}
\label{fig:F1_Spiral}
\end{figure}

The spiral (Swiss roll) dataset is a classification problem where the classifier separates data points from two spiral geometries (see \cref{fig:Fig1_Supp_Spiral}).
We see a promising advantage of higher-order methods in the experiments using traditional numerical methods shown in   \cref{fig:Fig3_Supp_Spiral}. 
To help reduce the number of learning epochs, we use a cyclic learning rate schedule  \citep{cycliclearningrate} to improve the chances of navigating local minima. The learning rate schedule is shown in \cref{fig:Fig2_Supp_Spiral}.
For the IMEX methods, a deeper network with $8$ hidden layers and a width of $8$ neurons is used with the SiLU activation function.
1000 data points are divided into ten batches. 
The training is carried out for 1000 epochs, and the results are visualized in \Cref{fig:F1_Spiral}. The IMEX methods have a computational advantage over the traditional \textsc{SGD} and RK methods, with the second-order Trapezoidal method slightly outperforming the first-order \textsc{Adam} as the training progresses.

\subsection{MNIST}\label{MNIST_Exp}

MNIST is a standard benchmark dataset for evaluating classification algorithms.
It consists of $60000$ training samples of handwritten digits and $10000$ samples for testing.
We reduced the training and testing sizes to $6000$ and $1000$, respectively, for faster training with 500 epochs.
The model used for classification has two hidden layers with 256 and 64 neurons, respectively. 

\begin{figure}[h]
\centering
\begin{subfigure}[b]{0.32\textwidth}
\includegraphics[width=\textwidth]{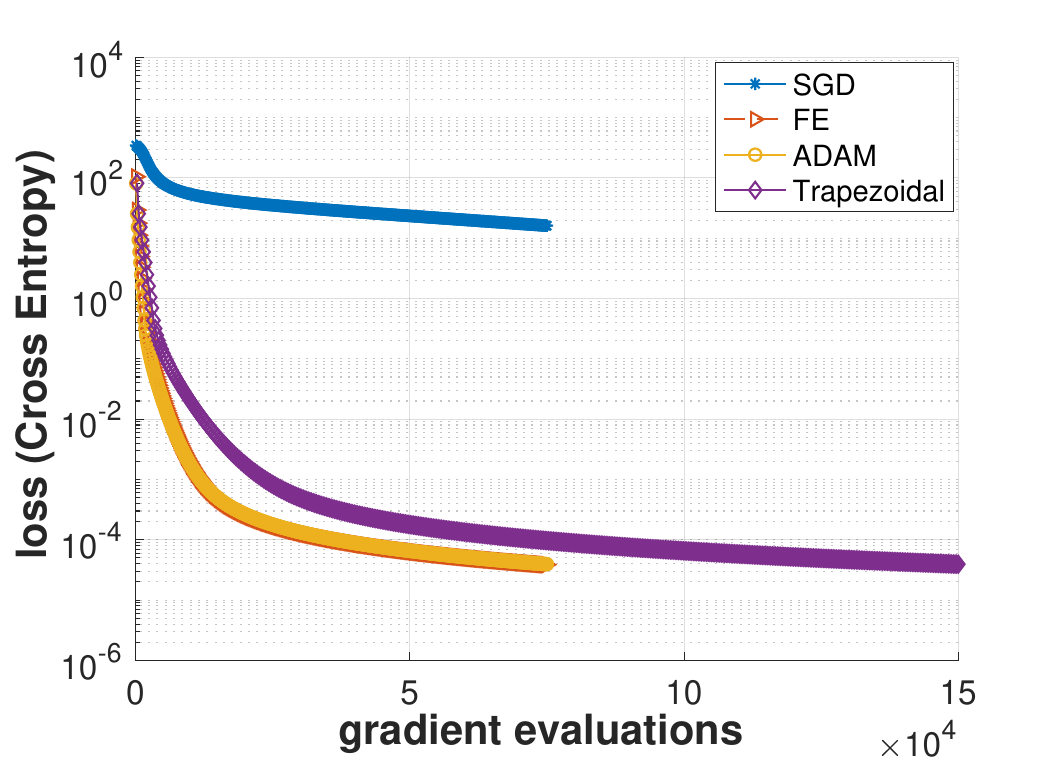}
\caption{lr = $0.001$, $\beta_1 = 0.9$, $\beta_2 = 0.999$,\\ batch = $150$}
\label{fig:Fig1_MNIST}
\end{subfigure}
\begin{subfigure}[b]{0.32\textwidth}
\includegraphics[width=\textwidth]{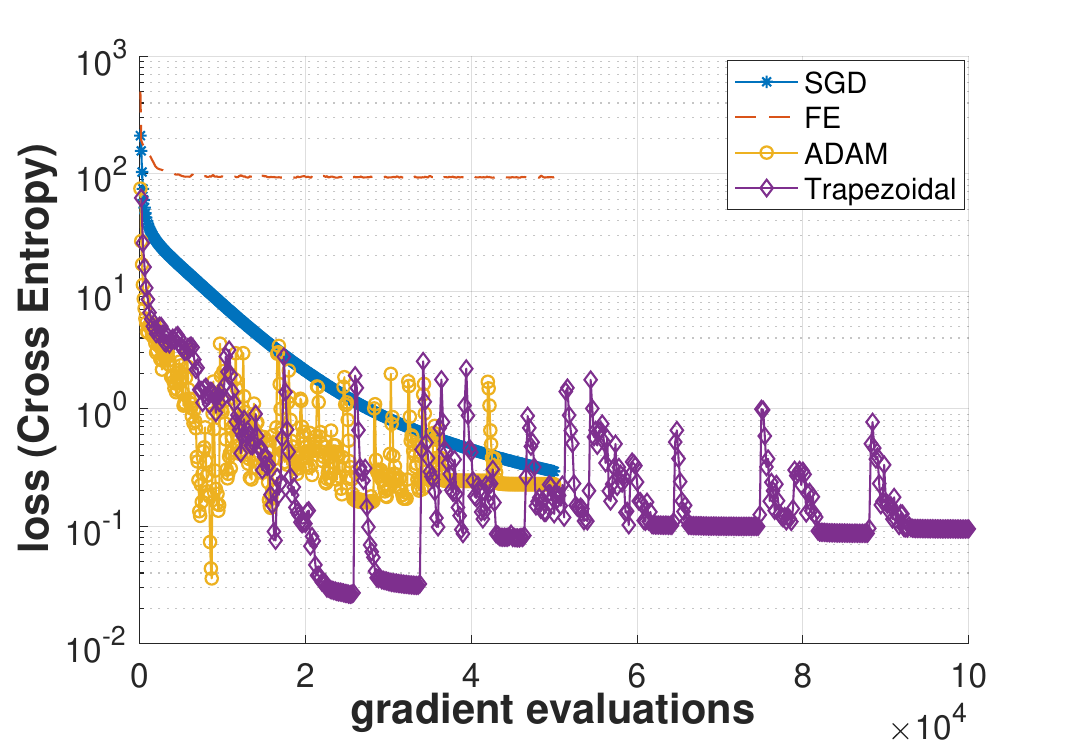}
\caption{lr = $0.01$, $\beta_1 = 0.9$, $\beta_2 = 0.999$,\\ batch = $100$}
\label{fig:Fig2_MNIST}
\end{subfigure}
\begin{subfigure}[b]{0.32\textwidth}
\includegraphics[width=\textwidth]{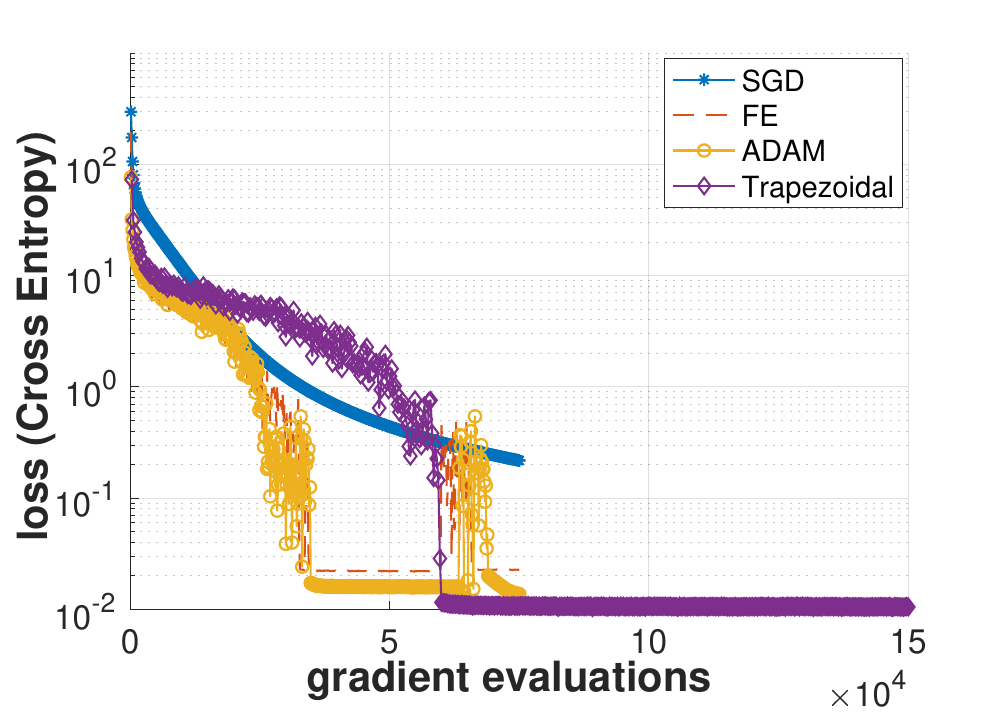}
\caption{lr = $0.01$, $\beta_1 = 0.9$, $\beta_2 = 0.99$,\\ batch = $150$}
\label{fig:Fig3_MNIST}
\end{subfigure}
\vspace{.3in}
\caption{Classification on MNIST Dataset. The hyper-parameters are reported for reference. \Cref{fig:Fig1_MNIST} denotes a stable learning trajectory (thanks to hyperparameter tuning), and we see IMEX methods are better than standard \textsc{SGD}. However, the second order method performs very similarly to the original first-order IMEX \textsc{Adam}, and we see no advantage of a higher order method in this setup. }
\label{fig:MNIST}
\end{figure}

The learning curves for training this model with different optimization methods are shown in \Cref{fig:MNIST}.
The new Trapezoidal method performs slightly worse than \textsc{Adam} but better than \textsc{SGD} (\cref{fig:Fig1_MNIST}).
However, it is interesting to note that the adaptive methods do well when the learning rate is large, or when the hyperparameters are not carefully ``tuned " to the best. The Trapezoidal method shows some advantages here (\cref{fig:Fig3_MNIST}).
The accuracy of the \textsc{Adam} and Trapezoidal methods in \cref{fig:Fig1_MNIST} are $94.7 \%$ and $94.7 \%$ respectively. In \cref{fig:Fig2_MNIST} it is $94.27 \%$ and $94.45 \%$, and in \cref{fig:Fig3_MNIST} the accuracy of \textsc{Adam} and Trapezoidal  methods are $94.56 \%$ and $94.43 \%$ respectively.

\subsection{CIFAR10 with VGG}
\begin{figure}[h]
\centering
\begin{subfigure}[b]{0.44\textwidth}
\includegraphics[width=\textwidth]{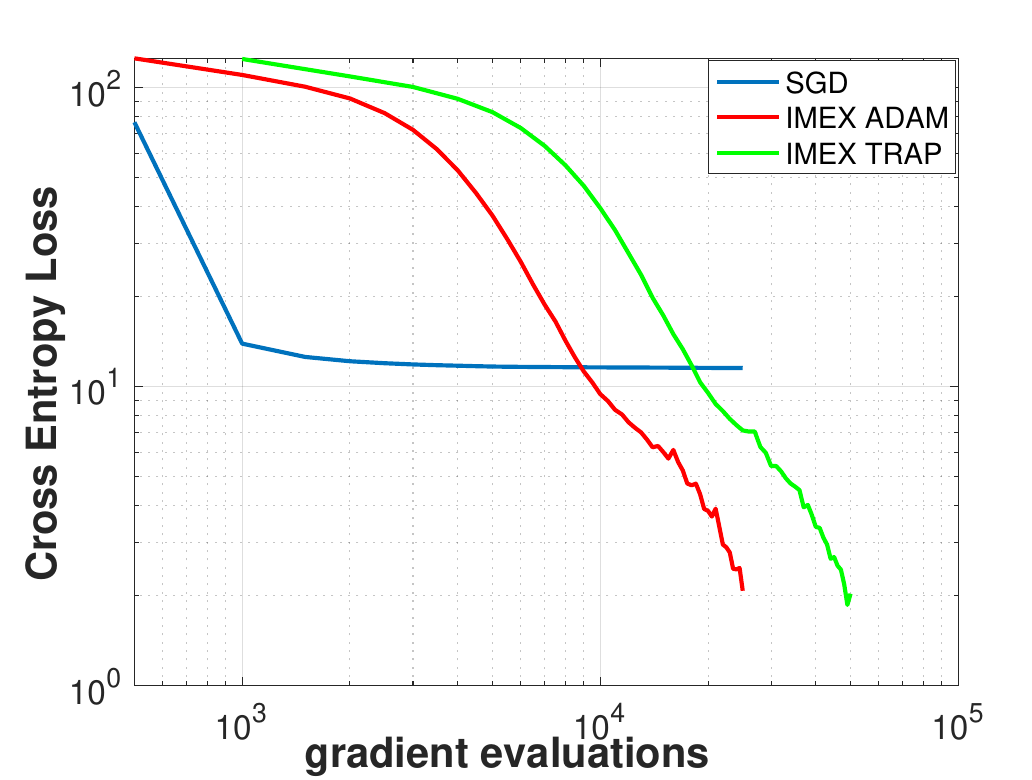}
\caption{Training loss}
\label{fig:Fig1_VGG16}
\end{subfigure}
\begin{subfigure}[b]{0.44\textwidth}
\includegraphics[width=\textwidth]{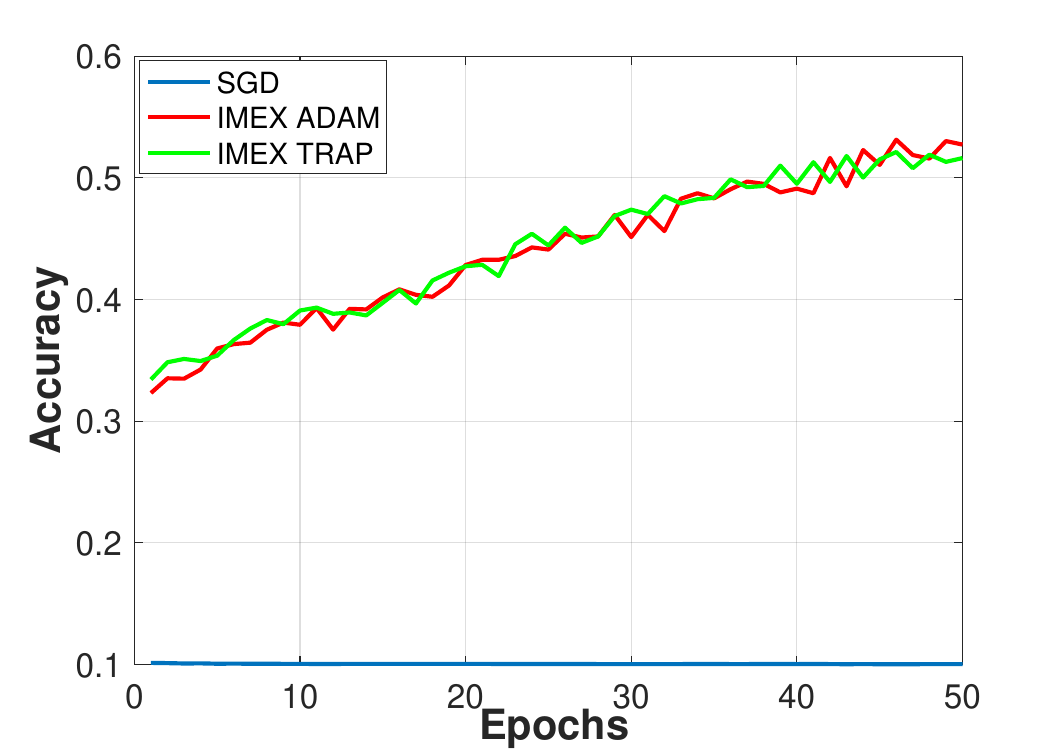}
\caption{Accuracy on hold-out data}
\label{fig:Fig2_VGG16}
\end{subfigure}
\vspace{.3in}
\caption{Classification task on the Cifar10 data using VGG architecture (lr = $0.001$, $\beta_1 = 0.9$, $\beta_2 = 0.999$, batches = $100$). The scaled training loss doesn't seem to give us any advantage using the new trapezoidal method. The accuracy is similar (refer \ref{fig:Fig2_VGG16}). Both the IMEX methods are better than \textsc{SGD}. \textsc{SGD} might need momentum or regularization techniques such as dropout, ridge or lasso regression.}
\label{fig:VGG16}
\end{figure}

Convolutional Neural Networks (CNNs) are ubiquitous in image classification tasks.
VGG-16 \citep{VGG} is a deep CNN architecture that won the ImageNet challenge prize in 2014. \Cref{fig:Fig1_VGG16}) shows the result of training with the CIFAR10 dataset using this architecture. Although the IMEX methods outperform \textsc{SGD},  
the new IMEX Trapezoidal method doesn't give any advantage during training for the same number of gradient evaluations. All three methods show about the same accuracy on the  test dataset (refer to \Cref{fig:Fig2_VGG16})  

\section{Conclusion}
\label{sec:conclusion}

This work extends the stochastic optimization method \textsc{Adam} \citep{kingma2017Adam}, widely used for neural network training, to an entire family of algorithms. The starting point of the analysis is the underlying \textsc{Adam} ODE \cref{eq:ODE}, to which the discrete \textsc{Adam} algorithm converges in the limit of infinitesimally small learning rates.

We establish that the standard \textsc{Adam} algorithm is the IMEX Euler discretization of the underlying \textsc{Adam} ODE \cref{eq:ODE}.
Based on this insight, we construct the new family of optimizers by applying a variety of GARK IMEX schemes to solvethe optimization ODE \cref{eq:ODE}. 
Employing time-stepping algorithms with orders of accuracy higher than one can result in optimization methods with faster convergence for neural network training.

Based on a variety of numerical experiments carried out, we select the second order IMEX Trapezoidal \textsc{Adam} as the method of choice, as it strikes a good balance between performance (decrease in the loss) and cost  (number of gradient evaluations per step).
The new IMEX Trapezoidal method outperforms standard \textsc{Adam} in the Lorenz63 experiment (\cref{subsec:L63_Exp}), sum of Gaussians (\cref{subsec:G3_Exp}) and Swiss roll classification problem (\cref{subsec:Spiral_Exp}) with deep architectures.
The second order method also works better with larger learning rates where other methods fail to converge or show oscillation in their loss value during training.

%
Several related directions are of interest to the authors for future research. Our experiments showed little computational advantage for the new IMEX methods over classical \textsc{Adam} on a number of datasets, including for MNIST and Cifar10.
Further study is required to evaluate when higher-order methods are needed in such cases.
Looking beyond IMEX GARK schemes, Linear Multistep Methods are another class of numerical time-stepping methods that can be applied in an IMEX fashion to partitioned ODEs. Linear Multistep Methods have a natural memory built in as they keep past information in the form of gradient evaluations or parameter values. Further research is needed to evaluate their potential in solving optimization ODEs with possible applications to training neural networks. How these methods compare against that of already established momentum methods \citep{NesterovMomentum, NesterovODE}, is a future endeavour for the authors. 


\acknowledgements

This research was supported by the Computational Science Laboratory  at Virginia Tech partly by Rolls Royce North American Technologies Inc., DOE ASCR DE-SC0021313, NSF DMS-2436357 and DMS-2411069. \\
Arash Sarshar's work used Jetstream2 at Indiana University through allocation CIS230277 from the Advanced Cyberinfrastructure Coordination Ecosystem: Services \& Support (ACCESS) program, which is supported by National Science Foundation grants \#2138259, \#2138286, \#2138307, \#2137603, and \#2138296.

\appendix

\section{ The \textsc{ADAM} optimizer} \label{app:original_adam}

The original \textsc{Adam} is listed below, where $\theta$s are the learnable parameters, $m$ is the momentum that captures the past gradients, and $v$ is the velocity which captures the square of the past gradients \cite{kingma2017Adam}.
The original literature has a bias correction term to correct for the zero initialization of the exponential moving average. That correction is not included here. 

	 	
	

\begin{algorithm}[H]
    \SetAlgoLined
    \KwData{Stepsize $h$; $\beta_1, \beta_2 \in [0,1)$, Loss function $\mathcal{L}(t,\theta)$ }
    \KwResult{Optimal parameters $\theta^*$}
    initialize $m_0 \gets 0, v_0 \gets 0, n \gets 0$\;
    \While{$\theta_n$ not converged}{
        $g_{n+1} \gets \nabla_{\theta} \mathcal{L}(t_{n+1}; \theta_n)$ \quad Calculate current gradient\;
        $m_{n+1} \gets \beta_1 \circ m_n + (1-\beta_1) \circ g_{n+1}$ \quad Update momentum\;
        $v_{n+1} \gets \beta_2 \circ v_n + (1-\beta_2) \circ g_{n+1}^{\circ 2}$ \quad update velocity\;
        $\theta_{n+1} \gets \theta_n - h \circ m_{n+1}/(\sqrt[\circ]{v_{n+1}} + \epsilon)$ \quad update parameters\;
        $n \gets n+1$.
    }
\caption{The \textsc{Adam} algorithm for solving stochastic optimization problems. Some nominal values are $h = 0.001$, $\beta_1 = 0.9, \beta_2 = 0.999$, $\epsilon = 10^{-8}$.}\label{alg:Adam}
\end{algorithm}

\section{Stability Plots}\label{app:stability_plots}
\begin{figure}[!ht]
\centering
\includegraphics[width=0.5\textwidth]{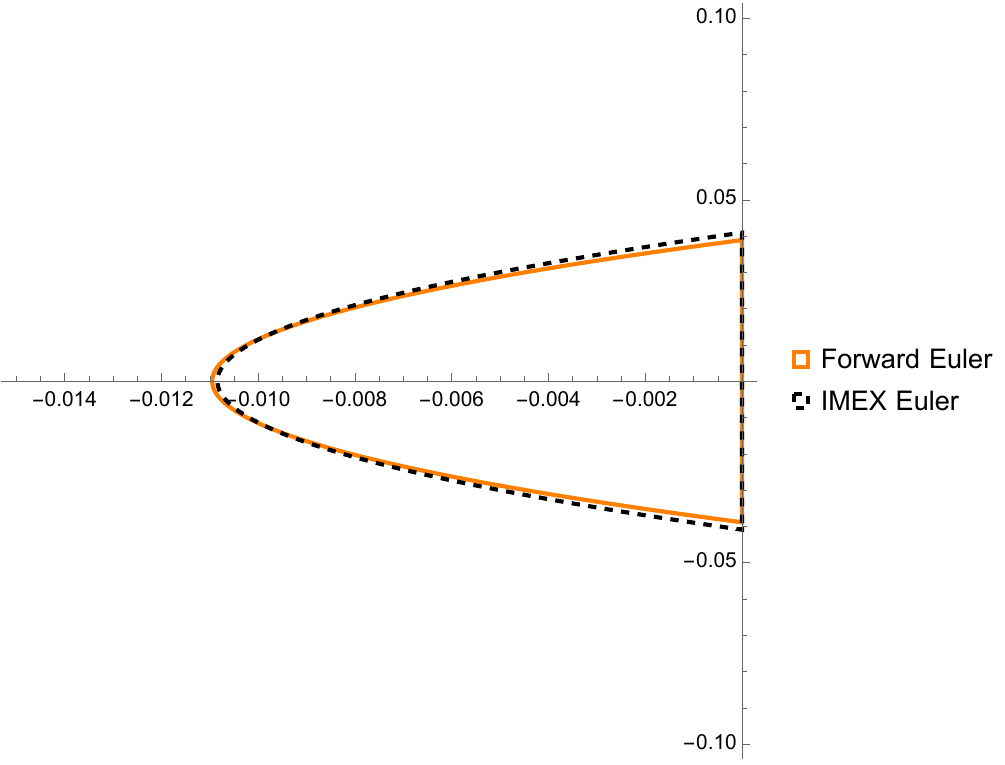}
\caption{Stability Region of the Linearized systems in \cref{eq:ForwardEulerODE,eq:Adam-partitioned-ODE}}
\label{fig:supp_stability_fe_vs_inmex}
\end{figure}

Continuing from \Cref{subsec:linear_stability_FE} and \Cref{subsec:linear_stability_imex_euler}, we plot the stability regions (refer to \cref{fig:supp_stability_fe_vs_inmex}) for the Forward Euler and IMEX Euler discretizations of the underlying ODE \cref{eq:ODE}. We fix the parameters to one set of values, i.e., $h = 0.0001, d = r = 1053.6, p = q = 10.005$.  The readers interested in the evolution of the stability region for other combinations of parameters are encouraged to investigate the Mathematica script\footnote{https://www.wolframcloud.com/obj/arashsarshar/Published/Adam-IMEX-Eigenvalues.nb} used to create these plots.

\section{Higher Order IMEX Methods}\label{app:higher_order_methods}

This section introduces two more higher-order methods developed for the underlying ODE in \Cref{sec:proposed_methodology} and satisfying the order conditions in \citep{GARKSandu}. As mentioned earlier, a time integration method can be characterized entirely by its Butcher tableaus, and we proceed to provide the parameterized tableaus for the methods. 

\begin{equation}
\label{eq:SSPRK3_LobattoIIIC}
\renewcommand{\arraystretch}{1.25}
\begin{tabular}{lc|ccc|ccc}
\multirow{3}{*}{Explicit} & $0$           & $0$           & $0$           & $0$           & $0$           & $0$            & $0$           \\
                          & $1$           & $1$           & $0$           & $0$           & $1$           & $0$            & $0$           \\
                          & $\frac{1}{2}$ & $\frac{1}{4}$ & $\frac{1}{4}$ & $0$           & $0$           & $\frac{1}{2}$  & $0$           \\ \hline
\multirow{3}{*}{Implicit} & $0$           & $0$           & $0$           & $0$           & $\frac{1}{6}$ & $\frac{-1}{3}$ & $\frac{1}{6}$ \\
 & $\frac{1}{2}$ & $\frac{1}{2} - l_{22}$   & $l_{22}$ & $0$                        & $\frac{1}{6}$ & $\frac{5}{12}$ & $\frac{-1}{12}$ \\
 & $1$           & $-1 + 8 l_{22} + l_{32}$ & $l_{32}$ & $2  - 8 l_{22}  -2 l_{32}$ & $\frac{1}{6}$ & $\frac{2}{3}$  & $\frac{1}{6}$   \\ \hline
                          &               & $\frac{1}{6}$ & $\frac{1}{6}$ & $\frac{2}{3}$ & $\frac{1}{6}$ & $\frac{2}{3}$  & $\frac{1}{6}$
\end{tabular}
\renewcommand{\arraystretch}{1.}
\end{equation}

The tableau (\cref{eq:SSPRK3_LobattoIIIC}) incorporates a third order SSP method \citep{SSP1,SSP2} for updating the explicit part of the underlying IMEX formulation and a \texttt{LobattoIIIC} methods for updating the implicit part \citep{Hairer1,Hairer2}. The coupling coefficients are derived from satisfying underlying order conditions. \texttt{LobattoIIIC} are both L-stable and has algebraic stability. $l_{22}$ and $l_{32}$ are parameters that can be tuned to get different methods. This method will be called "\texttt{SSPRK3LOBATTOIIIC}" in the experiment. 

\begin{equation}
\label{eq:RK4_LobattoIIIC}
\renewcommand{\arraystretch}{1.25}
\begin{tabular}{lc|cccc|cccc}
\multirow{4}{*}{Explicit} & \multicolumn{1}{l|}{$0$}           & $0$           & $0$           & $0$ & $0$ & $0$           & $0$           & $0$ & $0$ \\
                          & \multicolumn{1}{l|}{$\frac{1}{2}$} & $\frac{1}{2}$ & $0$           & $0$ & $0$ & $\frac{1}{2}$ & $0$           & $0$ & $0$ \\
                          & \multicolumn{1}{l|}{$\frac{1}{2}$} & $0$           & $\frac{1}{2}$ & $0$ & $0$ & $0$           & $\frac{1}{2}$ & $0$ & $0$ \\
                          & \multicolumn{1}{l|}{$1$}           & $0$           & $0$           & $1$ & $0$ & $0$           & $1$           & $0$ & $0$ \\ \hline
\multirow{3}{*}{Implicit} &
  \multicolumn{1}{l|}{$0$} &
  $0$ &
  $0$ &
  $0$ &
  $0$ &
  $\frac{1}{6}$ &
  $\frac{-1}{3}$ &
  $\frac{1}{6}$ &
   \\
 &
  \multicolumn{1}{l|}{$\frac{1}{2}$} &
  $\frac{1}{4}$ &
  $\frac{1}{4}$ &
  $0$ &
  $0$ &
  $\frac{1}{6}$ &
  $\frac{5}{12}$ &
  $\frac{-1}{12}$ &
   \\
 &
  \multicolumn{1}{l|}{$1$} &
  $0$ &
  $\alpha$ &
  $1-\alpha$ &
  $0$ &
  $\frac{1}{6}$ &
  $\frac{2}{3}$ &
  $\frac{1}{6}$ &
   \\ \hline
\multicolumn{1}{l}{} &
   &
  $\frac{1}{6}$ &
  $\frac{1}{3}$ &
  $\frac{1}{3}$ &
  $\frac{1}{6}$ &
  $\frac{1}{6}$ &
  $\frac{2}{3}$ &
  $\frac{1}{6}$ &
  
\end{tabular}
\renewcommand{\arraystretch}{1.}
\end{equation}

The tableau (\cref{eq:RK4_LobattoIIIC}) applies an RK4 method for the explicit part and a \texttt{LobattoIIIC} for the implicit part. The coupling coefficients are derived from satisfying the order conditions. It is to be noted that several new methods can be explored through the IMEX formulation. We enlisted just two that were tried out. $\alpha$ is the parameter that can be tuned. This method will be called "\texttt{RK4LOBATTOIIIC}" in the experiment.

\begin{figure}[!ht]
\centering
\vspace{.3in}
\begin{subfigure}[b]{0.45\textwidth}
\includegraphics[width=\textwidth]{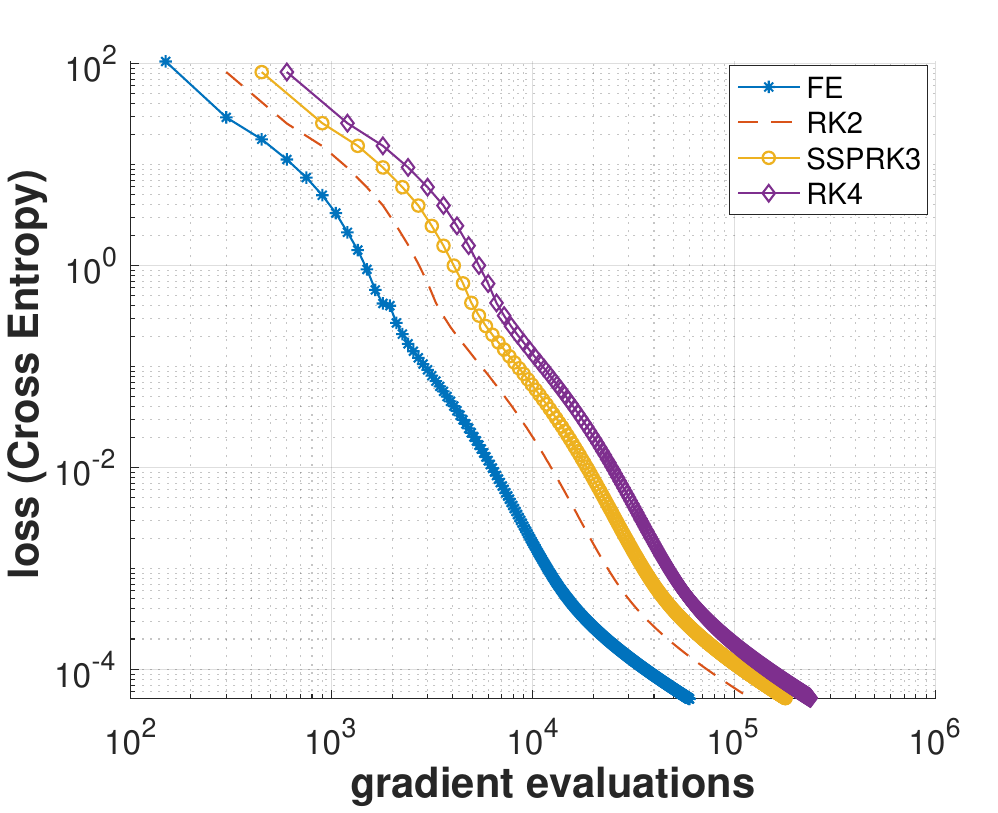}
\caption{lr = $1e-3$, $\beta_1 = 0.9, \beta_2 = 0.999$, \\ batches = $150$, hidden layer = ($256\times 64$)}
\label{fig:Fig1_Supp_MNIST}
\end{subfigure}
\begin{subfigure}[b]{0.45\textwidth}
\includegraphics[width=\textwidth]{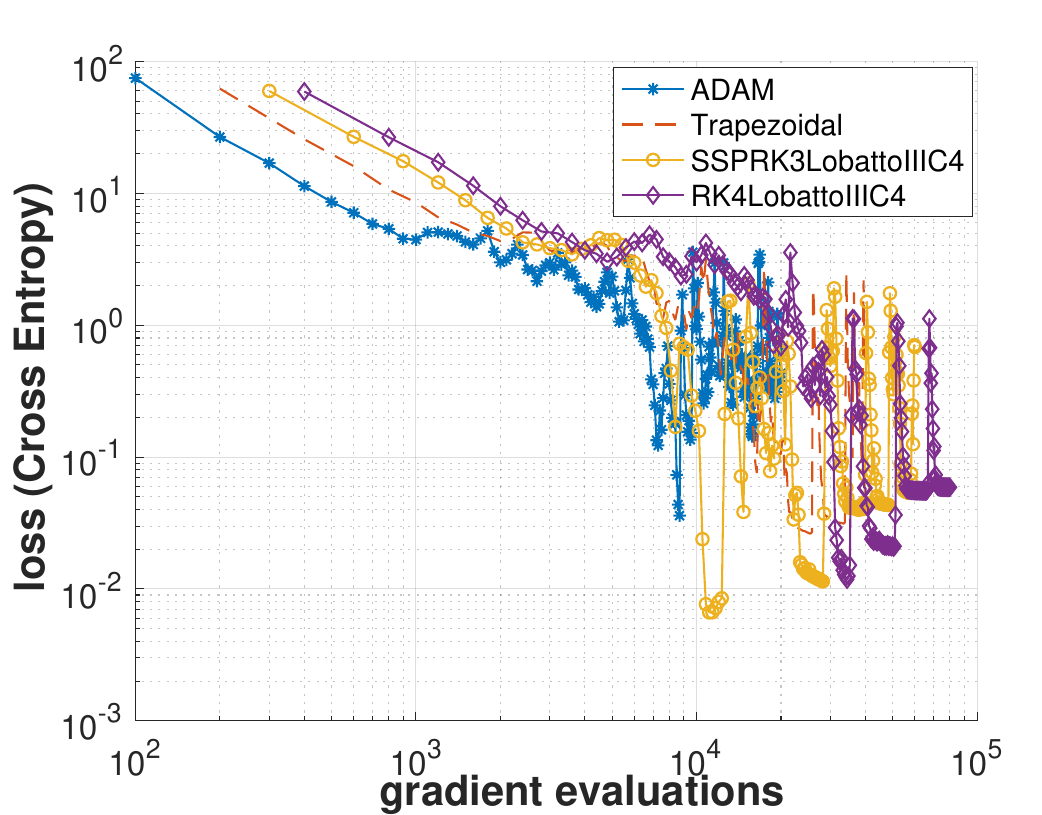}
\caption{lr = $1e-2$, $\beta_1 = 0.9, \beta_2 = 0.999$, \\ batches = $100$, hidden layer = ($256\times 64$)}
\label{fig:Fig2_Supp_MNIST}
\end{subfigure}
\vspace{.3in}
\caption{Classification on the MNIST dataset}
\label{fig:Supp_MNIST}
\end{figure}
A single experiment has been reported here using the newer higher-order methods compared to IMEX Euler and IMEX Trapezoidal methods (\cref{fig:Fig2_Supp_MNIST}).
The corresponding parameters are $l_{22} = 0.2, l_{32} = 0.1$ and $\alpha = 0.5$.
The user is free to tune these parameters as deemed fit.
One may also look at the stability plot for these methods (although there is no concrete proof that they will always work well!).
The batches are made lower with a higher learning rate to check the performance of the newer methods under unstable/unfavorable hyperparameters.
The newer higher-order methods go to a lower minima than the previous two methods. However, they are oscillatory due to their higher learning rate, which causes them to get stuck in a local minima.
Parameter tuning ($l_{22}, l_{32}, \alpha$) for the new higher order methods might work better and needs further investigation. 








\bibliographystyle{JMLSubmission/Bibliography_Style}
\bibliography{JMLSubmission/References}
\end{document}